\documentclass[12pt]{article}
\usepackage{graphicx,psfrag}
\usepackage{float}
\usepackage{amsmath, amssymb, graphics, setspace, relsize}

\newcommand \bra[1]{\left< {#1} \,\right\vert}
\newcommand \ket[1]{\left\vert\, {#1} \, \right>}
\newcommand \braket[2]{\hbox{$\left< {#1} \,\vrule\, {#2} \right>$}}
\newcommand{\bea}{\begin{eqnarray}}
\newcommand{\eea}{\end{eqnarray}}
\newcommand{\simgt}{\hbox{ \raise3pt\hbox to 0pt{$>$}\raise-3pt\hbox{$\sim$} }}
\newcommand{\simlt}{\hbox{ \raise3pt\hbox to 0pt{$<$}\raise-3pt\hbox{$\sim$} }}

\newcommand{\clfn}{\setcounter{footnote}{0}}

\newcommand{\LQ}{\Lambda_{\rm QCD}}
\newcommand{\alfs}{\alpha_{s}}
\newcommand{\msbar}{$\overline{\rm MS}$}

\begin{document}

\begin{titlepage}

    \begin{flushright}
      \normalsize TU-976\\
      \today
    \end{flushright}

\vskip1.5cm
\begin{center}
\Large\bf\boldmath
A Modern View of Perturbative QCD and \\
Application to Heavy Quarkonium Systems\footnote{
Lecture given at ``QCD Club," on 12 June 2014 at Univ.\ of Tokyo, Japan.
}
\unboldmath
\end{center}

\vspace*{0.8cm}
\begin{center}

{\sc Y. Sumino}\\[5mm]
  {\small\it Department of Physics, Tohoku University}\\[0.1cm]
  {\small\it Sendai, 980-8578 Japan}

\end{center}

\vspace*{0.8cm}
\begin{abstract}
\noindent
Perturbative QCD has made significant progress over the last few
decades.
In the first part, we present an introductory
overview of perturbative QCD as seen from a modern viewpoint.
We explain the relation between
purely perturbative predictions and predictions based on
Wilsonian effective field theories.
We also review progress of modern computational technologies and 
discuss 
intersection with frontiers of mathematics.
Analyses of singularities in Feynman diagrams play key roles
towards developing
a unified view.
In the second part, we discuss application of perturbative QCD, 
based on the formulation given in the first part, 
to heavy
quarkonium systems and the interquark force between static color charges.
We elucidate impacts on order $\LQ$ physics in the quark mass and
interquark force, which used to be considered inaccessible by perturbative
QCD.

\vspace*{0.8cm}
\noindent

\end{abstract}

\vfil
\end{titlepage}

\tableofcontents
\newpage

\section{Introduction}

More than 40 years have passed since the birth of QCD.
Accordingly there already exists plenty of accumulated knowledge
in perturbative QCD
owing to its history.
Let us quote a note written
by the late Kodaira in 2005 \cite{KodairaNote}, 
which summarizes the history of perturbative QCD 
very briefly:~
In the first decade from the mid-1970s,
correctness of perturbative QCD was confirmed qualitatively,
whereas quantitative predictions were found to  be difficult
due to existence of very large radiative corrections;
the second decade from the mid-80s was a time of contemplation,
during which the research field was subdivided and specialized,
and people went deeply into difficult problems;
in the third decade from the mid-90s, solutions to these problems were found
and perturbative QCD was progressively formulated
as precision science, where
predictions for observables in high-energy physics with order
10\% accuracy were becoming available.

Another decade has passed since then.
In the meantime, the accuracy as precision science improved further,
and many new predictions appeared, which were not possible before. 
Broadly speaking, today one sees a high degree of sophistication
achieved in every subdivided field 
of perturbative QCD.
One finds very interesting mature researches in each
of the subjects,
such as, jet physics, $B$ physics, physics of deep inelastic scattering,
top quark physics, quarkonium physics, etc.
In addition there are developments which are
common to or interconnect different fields of perturbative QCD, mainly in 
computational technologies.
There are excellent reviews on these subjects
\cite{Mueller:1989hs,Ellis:1991qj,Collins:2011zzd,Manohar:2000dt,Brambilla:2004jw,Smirnov:2004ym}.

The first part of this 
lecture is intended to present an overivew of perturbative
QCD to non-experts, including graduate course students.
Given the current status of being subdivided into specialized
fields,
it is not easy for most people to obtain a good perspective over
the whole of perturbative QCD.
Nevertheless, we find that there is a flow in  
recent developments, 
which may become a useful notion
for developing a unified view of perturbative
QCD across various subdivided fields.
In this lecture, we raise attention on this flow
and try to provide a perspective from a unified viewpoint.
Although this is the aim, the author  
excuses in advance that the contents are much influenced by
his own research career and would be biased, as
it is highly non-trivial
to understand the details of perturbative QCD in various
subjects, reconstruct them and extract the essence.
It should be noted that there are many important 
concepts and useful frameworks which are not covered
by this lecture.
In particular, it is a marked deficit
that this lecture barely covers developments of perturbative
QCD related to the LHC physics, which are actively 
evolving at this moment and many of which are yet to be
formulated in a well organized way.

We pay particular attention to 
singularities in Feynman amplitudes
as a key in formulating perturbative QCD.
Perturbative QCD is a theoretical tool 
for elucidating dynamics of QCD and giving quantitative
predictions to various phenomena of QCD.
Hence, our tasks  in perturbative QCD are to understand the nature
of its large radiative corrections and to give their
quantitative descriptions.
To realize the tasks,
the main theoretical
issues can be boiled down to the following two points:
(1) To develop a method on how to decompose and systematically organize
the radiative corrections, and
(2) to elucidate the nature of the radiative corrections
contained in the individual parts of the decomposition
(which are simplified by the decomposition).
Singularities in amplitudes play key roles in both of these
issues.

To realize (1), theoretical frameworks
such as factorization,
various effective field theories (EFTs) and opertator product expansion
(OPE) have been exploited.
In a modern language, we can develop these frameworks
utilizing singularities in Feynman amplitudes.
It is different from, as well as complementary to, 
the conventional methods which divide integral
regions in momentum space into separate domains.
(In this lecture we take Wilsonian EFTs and OPE as
representatives of these methods, although the factorization 
framework \cite{Collins:1989gx}, which is conceptually
close, deserves an extensive explanation on its own.)

To understand the nature of the
radiative corrections in (2),
there are on-going efforts in the areas 
covering both mathematics and
physics.
It has been known empirically that, in computations of
higher-order radiative corrections, final analytical results 
turn out to be surprisingly compact and simple, in comparison
to an enormous amount work required at intermediate stages.
While singularities of Feynman amplitudes
must be playing crucial roles behind this
empirical fact, there are unsolved questions towards
understanding the mechanism.
We review basic knowledge and some recent aspects.

In the latter part of this lecture, we discuss
an application of
perturbative QCD to heavy quarkonium systems using
the above formulation.
Detailed properties of heavy quarkonium systems can be investigated both
qualitatively and quantitatively 
thanks to recent developments of perturbative QCD.
Although there exist a wide variety of
studies on heavy quarkonium \cite{Brambilla:2004wf,Brambilla:2010cs}, we do not present a
survey of those studies.
In this lecture we review the studies with respect to more
fundamental properties and show that the formulation of perturbative 
QCD works consistently.
In particular we clarify the relation between 
the intrinsic QCD scale ($\LQ\approx 300$~MeV) and
properties of the static QCD potential
at $r\simlt \LQ^{-1}$.
We also provide physical interpretations.

The lecture is organized as follows.
(See also the table of contents in page 1.)
In Sec.~2 we give a quick overview of perturbative QCD.
In Sec.~3 we present more details of the overview.
These sections deal with the formulation of perturbative QCD.
Its application to heavy quarkonium
systems is reviewed in Sec.~4.
Concluding remarks are given in Sec.~5.
We give a proof of a relation
concerning properties of the static potential
in the Appendix.

\section{Quick Overview of Perturbative QCD}
\clfn

\subsection{What is perturbative QCD?}

Today, when people refer to ``perturbative QCD predictions,'' 
one notices that
they can be classified into (at least) three different types.
One should be careful, since
without properly distinguishing between them,
one may be led to confusions and eventually to unwanted
errors.
These three types can be defined as follows:
\begin{itemize}
\item[(a)]
Prediction of an observable in the series expansion in $\alfs$.
\item[(b)]
Prediction of an observable in the framework of a Wilsonian 
EFT.
\item[(c)]
Prediction of an observable assisted by model predictions.
\end{itemize}

The prediction (a) is literally a perturbative expansion
of an observable
in the strong coupling constant $\alfs$.
The observable needs to be free of IR divergences.
Although the prediction is purely perturbative, it
is known to contain intrinsic
uncertainties in powers of $\LQ$ in the form $\sim (\LQ/E)^n$, where $n$ is
some positive integer dependent on the observable and $E(\gg \LQ)$
denotes a scale associated with the observable (e.g.\
the center-of-mass energy for a total cross section).

In the prediction (b), one performs OPE within a low energy EFT.
It gives a systematic expansion of an observable in powers of $\LQ/E$.
In this formulation, uncertainties in (a) is replaced by 
non-perturbative matrix elements.
One should not, for instance, 
add a non-perturbative quantity
of the prediction (b) to the prediction (a),
which is an error one could make without recognizing it.

The prediction (c) is less solid compared to (a) and (b).
While the main part of the prediction (c) is given by
a literal perturbative expansion of type (a),
the prediction also includes parts which are not
based on QCD.
Theoretical predictions of various observables 
in high-energy experiments (in particular
hadron collider experiments) depend on 
hadronization models and parton distribution functions (PDFs).
These are necessary ingredients in Monte Carlo (MC) simulations
used to compare theoretical predictions with
experimental data. 
While final states in perturbative QCD
include quarks and gluons, in real experiments we observe instead 
hadrons in final states, hence quarks and gluons need
to be hadronized in relating theoretical predictions
to experimental data.
Predictions of hadronization processes
rely on models such as string
hadronization models and cluster hadronization
models.
These models are, however, difficult to 
relate to QCD at the fundamental level.
On the other hand,
the distributions of quarks and gluons inside the initial
proton and/or antiproton are described by PDFs.
To predict PDFs we need an initial condition for the evolution
equation. 
At present it is difficult to predict the initial condition
from the first principle of QCD and they are replaced by
some ansatz or a prediction by a model.
Systematic uncertainties originating from these model
predictions are
difficult to control with high precision.
Typically order 10\% accuracy is achieved
for this type of predictions of observables in LHC experiments, 
although details depend
very much on observables.\footnote{
The difference of type (b) and type (c) predictions 
is that the former
is systematic while the latter is not in parameterizing
the part which are difficult to compute from the first
principles of QCD.
}

\subsection{Remarkable progress of computational technologies 
in the last 10-20 years}

Development of perturbative QCD in the last ten
to twenty years has been largely based on
remarkable progress of computational technologies 
during this period.
In this short overview we would like to emphasize especially developments
in the following fields of technologies:
\begin{itemize}
\item[(i)]
Higher-loop corrections\footnote{
Some examples of the higher-loop computations
we are concerned here are the 4-loop coefficient of the
beta function, 4-loop coefficient of the anomalous dimension,
4-loop correction to the $R$-ratio,
3-loop correction to the pole--{\msbar}
mass relation, 3-loop correction to the static QCD potential,
etc.
We should also include many higher-loop computations related to
LHC physics, such as the Higgs production cross section
at next-to-next-to-leading order (NNLO), where the
boundary between the technologies (i) and (ii) becomes
somewhat obscure.
}
\item[(ii)]
Lower-order (NLO/NLL) corrections to complicated processes
\item[(iii)]
Factorization of scales in loop corrections
\end{itemize}

Computations of higher-loop corrections [(i)]
can essentially
be regarded as processes of
resolution of singularities in multi-loop
integrals.
Both nummerical and analytical methods have been developed to
evaluate higher-loop corrections.
Concerning numerical methods, an essential problem has been
solved by application of the theorem on resolution of singularities
\cite{Bogner:2007cr}.
There exist certain algorithms (known as sector decomposition
\cite{Binoth:2000ps,Binoth:2003ak})
to resolve any type of singularities
in numerical loop computations by a finite number of steps.
Thus, we can achieve higher-loop computations as we increase
computational power that can be invested.
(There have also been many studies
on improving efficiencies of
algorithms for numerical evaluations.)
On the other hand, general algorithms have not been found yet
for analytic evaluation of
higher-loop corrections.
There have been many developments in this direction, and we
see rich activities
in the fields which overlap with frontiers of mathematics.

Strongly motivated by the LHC experiments, 
rapid progress is taking place in a variety of
technologies to compute 
lower-order [next-to-leading order (NLO) or
next-to-leading logarithmic (NLL)] corrections to complicated processes
[(ii)].
These technologies enable us to cope with 
proliferation of Feynman diagrams and existence of
many kinematical variables.
These technologies provide practical computational
tools, often in conjunction with MC event generators.
Nevertheless, at present these exploited
technologies still consist of a collection
of various separate techniques and 
a general or systematic theoretical formulation is not yet available.

Factorization of energy scales in a Feynman diagram
can be realized
by a method called asymptotic expansion or
integration by regions of a loop integral [(iii)]  \cite{Smirnov:2002pj}.
This technology provides a precise 
foundation for constructing Wilsonian EFTs.
It enables us to compute efficiently higher-order corrections
to Wilson coefficients in EFTs.
As a variety of EFTs have been formulated to tackle
different physics targets using perturbative QCD,
this technology has become an indispensable tool for computations.

A salient feature of all
the above technologies is that
they rely on dimensional regularization
as the common theoretical basis.
Perhaps the name ``dimensional regularization'' does not describe its 
characteristics appropriately; it  is 
essentially a regularization by analytic continuation of loop integrals.
This regularization method is
contrasting to, as well as complementary to, more
conventional cut-off regularization methods which
were primarily used in the early days of loop computations.
In particular, dimensional regularization permits
use of analyticity of loop integrals flexibly.

\subsection{Comment on Impacts on Physics Insights}

Let us breifly comment on impacts of the above developments of perturbative
QCD on physics insights, namely new interpretations, viewpoints
or concepts in physics which resulted from the developments.
We find that,
to date, these impacts are scattered over various specific fields
within perturbative QCD. 
Indeed, the developments of perturbative QCD revealed rich structures
in these specialized subjects individually.
On the other hand,
we fail to extract general concepts covering all or most of them,
and we are
yet to frame a general overview.

One may find some examples in the following subjects.
Construction of various EFTs and accurate predictions based
on them
triggered new paradigms, 
such as heavy quark effective theory (HQET) for $b$-physics, 
or soft-collinear effective theory (SCET) for jet physics.
Another example is an impact on ${\cal O}(\LQ)$
physics in the heavy quark mass and interquark force.
Apparently there exists a belief that powers of
$\LQ=\mu \exp [-2\pi/(\beta_0\alfs)]$
cannot appear in perturbative expansions {\it at all},
since the expansion of $\LQ$ in $\alfs$ vanishes
to all orders.
Recent studies show, however, that this is
not necessarily the case.
We will explain the latter example in Sec.~4.

\section{Overview of Perturbative QCD in More Detail}
\clfn 

In this section
we explain further details of the aspects
of perturbative QCD explained in Secs.~2.1 and 2.2.
Nonetheless, we do {\it not} cover the subjects
in (c) of Sec.~2.1 and (ii) of 
Sec.~2.2, 
since we consider that we have not yet
reached to a stage 
to discuss theoretical formulation
of these subjects concisely.

\subsection{\boldmath Perturbative expansion in $\alfs$}
\label{Sec:PurelyPTQCD}

Perturbative QCD is a theory of quarks and gluons
based on the QCD Lagrangian
${\cal L}_{\rm QCD}(\alfs,m_u,m_d,m_s,\dots;\mu)$,
where $m_q$ denotes the mass of the quark $q$ and
$\mu$ denotes the renormalization scale.
A perturbative prediction of an observable is given as a series
expansion in $\alfs$, whose expansion coefficients
are given as functions of $m_q$, $\mu$ and kinematical variables.
The perturbative prediction has the same input parameters as the full QCD.
Usually the renormalization scale $\mu$ is not
considered as an input parameter but rather a parameter
used to estimate stability and uncertainties
of the prediction.
The prediction is systematic in the sense that it
has its own way of estimating errors
(without comparing to the corresponding experimental value), by using the
$\mu$ dependence as well as by examining convergence
of the series expansion.
In this regard, perturbative QCD differs from models,
which generally do not have systematic ways to estimate
errors and have more input parameters than the full QCD.
\begin{figure}[t]
\begin{center}
\includegraphics[width=8cm]{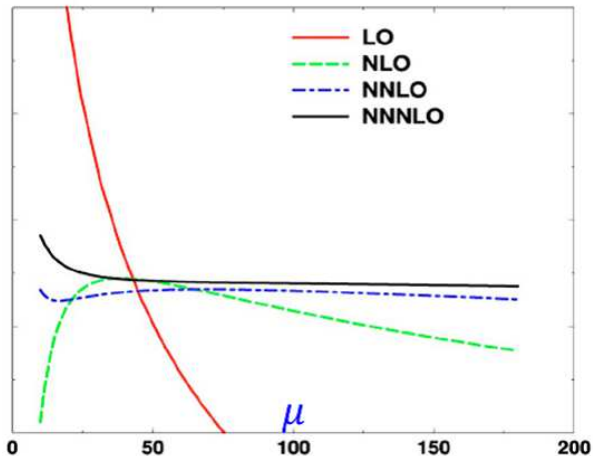}
\end{center}
\vspace*{-.5cm}
\caption{\small
Plot showing schematically renormalization scale ($\mu$)
dependence of a prediction of a physical observable
in perturbative QCD.
\label{mu-dep}}
\end{figure}
\noindent
Fig.~\ref{mu-dep} shows schematically typical $\mu$ dependence
of a perturbative prediction as we increase the order
of the expansion.
Since the all-order prediction is formally independent
of $\mu$, we observe that the $\mu$ dependence
decreases as we include higher-order terms.

There are two classes of observables which are 
considered to be predictable
by perturbative expansions in $\alfs$.
The first class consists of inclusive observables
with respect to final-state hadrons.
The most well-known example is the
$R$-ratio:
\bea
R(E)\equiv \frac{\sigma(e^+e^- \to {\rm hadrons};E)}
{\sigma(e^+e^-\to \mu^+\mu^-;E)}
=\sum_q 3Q_q^2\left[
1+\sum_{n=1}^\infty c_n(E/\mu)\,\alfs^n
\right] .
\label{R-ratio}
\eea
Other observables in this class include decay widths
of a non-colored particle (such as the tau lepton, 
weak boson, Higgs boson, etc.) or
distributions of non-colored particles
(such as lepton energy distribution)
after integrating over the phase space of final-state hadrons.
In perturbative QCD we compute these observables with respect to
final-state partons (quarks and gluons)
instead of hadrons, integrating over their phase space.
We compare these observables 
with the experimental values assuming that effects
of hadronization can be neglected.
Logically this corresponds to assuming
that the complete Fock space spanned by
hadron states is equivalent to the space spanned by
the states of free partons.\footnote{
This is a highly non-trivial assumption
under the existence of color confinement and spontaneous
chiral-symmetry breakdown in QCD.
}
The identification of inclusive observables for
hadronic and partonic final states
is a testable hypothesis, by comparing
predictions to
experimental data.
So far this hypothesis seems to work reasonably well.

The second class of observables are
those of heavy quarkonium states.
The heavy quarkonium states are the only known hadronic states, whose individual
properties can be predicted using perturbative QCD, due to the
heavy mass of quarks and asymptotic freedom of QCD.
Specifically we can compute their energy spectrum, decay width,
leptonic branching ratios,\footnote{
These can be translated to production 
cross sections in $e^+e^-$ collisions.
} and transitions rates.
We will discuss the energy levels of
heavy quarkonium states in Sec.~4.

As already mentioned, a prediction given as a perturbative series
in $\alfs$ generally contains uncertainties
of order $(\LQ/E)^n$.
These uncertainties originate from 
IR sensitivities of higher-order corrections
and are referred to as contributions of ``IR renormalons''
\cite{Beneke:1998ui}.
We explain how the uncertainties arise, 
taking the $R$-ratio as an example.\footnote{
The explanation given below is slightly sloppy and is correct
only qualitatively.
In a more solid argument,
one should carefully choose a gauge-independent formulation
(such as to choose the background gauge or to use the ``large-$\beta_0$
approximation'' \cite{Beneke:1994qe}).
Quantitatively it also matters how to fix the non-logarithmic
term of the one-loop gluon self-energy.
}

\begin{figure}
\begin{center}
\includegraphics[width=16cm]{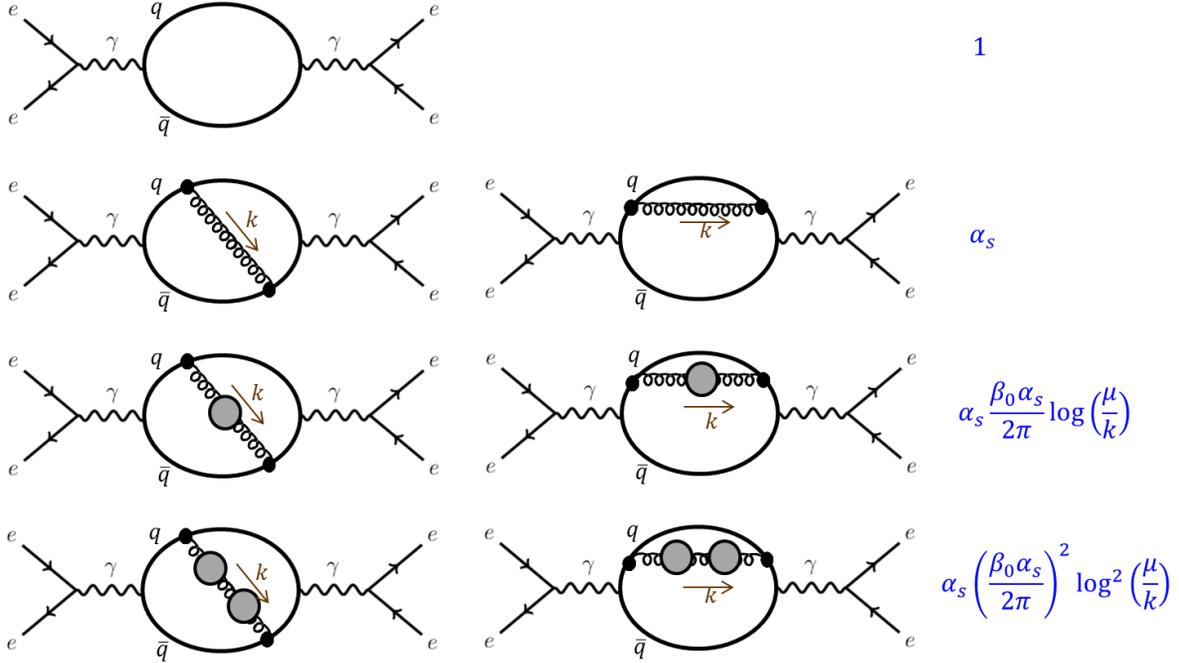}
\end{center}
\vspace*{-.5cm}
\caption{\small
Class of diagrams called bubble-chain diagrams contributing
to the forward  scattering amplitude for $e^+e^- \to e^+e^-$.
The bubble in the diagrams represents the sum of the 
one-loop self-energy diagrams of the gluon.
\label{bubble-chain}}
\end{figure}
According to the optical theorem, 
the $R$-ratio can be computed from the imaginary part of
the forward scattering amplitude for $e^+e^- \to e^+e^-$.
In Fig.~\ref{bubble-chain}, we show
a class of Feynman diagrams, known as 
bubble-chain diagrams,
relevant for this computation.
(The bubble in the diagrams represents the sum of the 
one-loop self-energy diagrams of the gluon.)
The displayed diagrams at ${\cal O}(\alfs^0)$ and ${\cal O}(\alfs^1)$ are the
only ones that contribute at these orders.
At ${\cal O}(\alfs^2)$ and beyond, the diagrams shown in the figure
are only part of the whole contributions.
As indicated in the figure, before 
integrating over the gluon momentum $k$, the contribution of
the bubble chain at ${\cal O}(\alfs^n)$ is given by
$\alfs(\mu) [\beta_0\alfs(\mu)\log(\mu/k)/(2\pi)]^{n-1}$.
($\beta_0$ denotes the coefficient of the one-loop beta function.)
Therefore, we can incorporate the effect of the resummation of 
the bubble-chain diagrams to all orders
by replacing the coupling constant $\alfs(\mu)$ in
the ${\cal O}(\alfs)$ diagrams by the one-loop running coupling
constant 
\bea
\alfs(k)=\frac{\alfs(\mu)}{1-\beta_0\alfs(\mu)\log(\mu/k)/(2\pi)}
= \frac{2\pi}{\beta_0\log(k/\LQ)}.
\label{1Lrunningcoupling}
\eea
The following characteristic feature of the running
coupling constant is relevant: 
it becomes large at IR and this occurs at the intrinsic QCD
scale $\LQ$.
In a similar manner,
the effect of the bubbule-chain diagrams at each order
can be incorporated by replacing the coupling constant $\alfs(\mu)$ in
the ${\cal O}(\alfs)$ diagrams by the factor
$\alfs(\mu) [\beta_0\alfs(\mu)\log(\mu/k)/(2\pi)]^{n-1}$.
\begin{figure}
\begin{center}
\includegraphics[width=7cm]{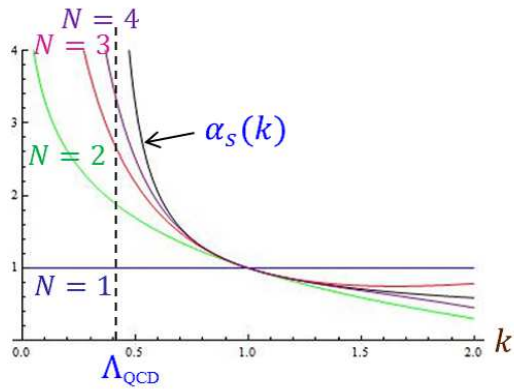}
\end{center}
\vspace*{-.5cm}
\caption{\small
Plot showing $k$ dependences of
the one-loop running coupling constant
$\alpha_s(k)$ [Eq.~(\ref{1Lrunningcoupling})]
and the sums of the leading logarithms 
$\sum_{n=1}^N \alfs(\mu) [\beta_0\alfs(\mu)\log(\mu/k)/(2\pi)]^{n-1}$
for $N=1$, 2, 3 and 4.
$\LQ$ represents the position of the pole of $\alpha_s(k)$.
\label{RunningCoupling}}
\end{figure}
Fig.~\ref{RunningCoupling} shows the summation of this factor up to 
${\cal O}(\alfs^N)$
for $N=1$, 2, 3 and 4.
One can see that as the higher-order diagrams
are incorporated the sum of the bubble-chain contributions
approaches the running coupling constant $\alfs(k)$.
This shows that, although at each order the scale $\LQ$ does not
appear explicitly, the higher-order corrections know about this intrinsic
scale and know that the strong interaction becomes strong
at this scale.

The consequences of incorporating these higher-order
diagrams are as follows.
If we estimate
the perturbative series in eq.~(\ref{R-ratio}) by these bubble-chain
contributions, the series diverges at high orders,
reflecting the increase of the higher-order contributions
in the IR region.\footnote{
The important point is that $\alfs(k)$ becomes large 
at $k\sim\LQ$ rather than it has a pole at this scale.
For instance, the series is similarly divergent
in the case that we incorporate the effects of
the two-loop coefficient of the beta function when its sign is
taken to be opposite to that of the one-loop coefficient.
In this case the running coupling has an IR fixed-point and
therefore is finite down to $k=0$.
}\pagebreak
\begin{figure}[h]
\begin{center}
\includegraphics[width=7cm]{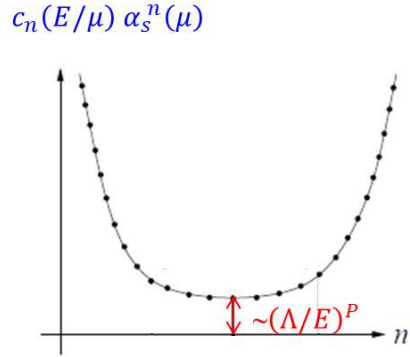}
\end{center}
\vspace*{-.5cm}
\caption{\small
Diagram showing typical $n$-dependence of a perturbative
series of an observable, based on renormalon estimates.
In the case of the $R$-ratio, $P=4$.
\label{asympt-series}}
\end{figure}
Therefore, the series is (at best) asymptotic and there is a
limitation in the achievable accuracy of the prediction
even if we know the perturbative series up to arbitrarily high orders.
Namely, until the order $n_0\approx 2\pi/(\beta_0\alfs)$, 
each term $c_n$ of the series
in eq.~(\ref{R-ratio}) becomes smaller; it takes a minimal value
$c_n \sim (\LQ/E)^P$
at $n= n_0$; for $n>n_0$, $c_n$ diverges rapidly.
See Fig.~\ref{asympt-series}.
The appearance of the positive power of $\LQ$ reflects
the fact that the higher-order terms know this intrinsic scale.\footnote{
Generally $P$ is a positive
integer, and in the case of the $R$-ratio, $P=4$.
} 
In view of general properties of asymptotic series,
we may expect that the sum of the series approaches the
true value for $n<n_0$, and that the series has at least
an error of order $(\LQ/E)^P$.\footnote{
A famous example is 
an asymptotic series 
\bea
n! = \sqrt{2\pi n} n^n e^{-n}\left(
1+ \frac{1}{12n}+\frac{1}{288n^2}-\frac{139}{51840n^3}
-\frac{571}{2488320n^4}+\cdots \right),
\eea 
whose leading term is known as Stirling's formula.
The difference of a truncated series and the true value
is of the order of 
the last term of the truncated series.
Hence, we can improve accuracy by including higher-order terms
until we reach the minimum term of the series.
}
It turns out that often such estimates based on bubble-chain
diagrams
are good estimates of the true corrections,
when the first several terms of the series
are known exactly.

\subsection{OPE in a Wilsonian effective field theory}

Consider a Wilsonian low-energy EFT, written in terms of light quarks
and gluons.
Formally it can be constructed from the full QCD by
integrating out high-energy modes above a factorization scale
\pagebreak
$\mu_f(\gg \LQ)$ in a path-integral formulation of the theory.
The Lagrangian of the EFT can be written in a form
\bea
{\cal L}_{\rm EFT}(\mu_f)=\sum_i g_i(\mu_f) \, {\cal O}_i
(\psi_q,\bar{\psi}_q,G_\mu),
\label{EFT-Lagrangian}
\eea
which is a sum of operators ${\cal O}_i$ composed of
light quarks and gluons, whose energies and momenta are restricted
to be below $\mu_f$.
The effective coupling constant $g_i(\mu_f)$ multiplying
each operator is called a Wilson coefficient, which is
determined such that the physics at $E<\mu_f$ is unchanged
from the full QCD.
This is expected to be possible by introducing infinitely
many operators ${\cal O}_i$ in the Lagrangian.
Since $g_i(\mu_f)$'s include only effects of UV degrees of
freedom ($E>\mu_f$), they can be computed reliably using perturbative
QCD.
In practice
there are two ways to determine the Wilson coefficients.
One way is to compute various $S$ matrix elements with
external momenta of order $E$, where
$\mu_f \simgt E \gg \LQ$,
in both EFT and full QCD in expansions in $\alfs$,
and to require that both computations give the same results.
This is known as a matching procedure.
The other method is to apply asymptotic expansion of
diagrams \cite{Smirnov:2002pj}, 
which determines the operators and Wilson coefficients
of EFT in an efficient way;
we will explain this latter method in Sec.~3.4.
The Wilson coefficients computed using perturbative QCD
should be free of 
uncertainties by IR renormalons,
since the region of integration (above $\mu_f$)
does not include the domain
where the strong coupling constant is large.
Thus, the EFT Lagrangian eq.~(\ref{EFT-Lagrangian})
consists of Wilson coefficients, which effectively
contain information on UV degrees of freedom, and
operators composed of dynamical variables representing
IR degrees of freedom.

We can perform an operator-product-expansion (OPE)
in the EFT, for an observable $A(P)$
which includes a high scale $P(\gg \mu_f)$.
An example of $P$ is a mass of a heavy particle, whose effects
enter $A(P)$ only through Wilson coefficients, since the
heavy particle has been integrated out and is absent as a dynamical field
of EFT.
Within the EFT, we can compute the observable
expressed in terms of Wilson coefficients and matrix elements of
operators.
Furthermore,
we can perform a multipole expansion of operators, which is an
expansion in terms of derivatives acting on the fields of the EFT:
\bea
&&
A(P) = g_1(\mu/P)\bra{n}{\cal O}(x)\ket{n}+
\frac{g_2(\mu/P)}{P^2}\,\bra{n}\partial_\alpha {\cal O}(x)
\partial^\alpha {\cal O}(x)\ket{n}
\nonumber\\ &&
~~~~~~~~
+
\frac{g_3(\mu/P)}{P^4}\,\bra{n}\partial_\alpha\partial_\beta {\cal O}(x)
\partial^\alpha \partial^\beta{\cal O}(x)\ket{n} + \cdots .
\label{OPE}
\eea
Here, ${\cal O}(x)$ represents symbolically local
operators without derivatives.
Each derivative operator $\partial$
corresponds to the energy-momentum ($\sim k$)
of the fields in ${\cal O}(x)$.
Hence, this expansion constitutes an
expansion in $k/P(\ll 1)$, where $k$
is restricted to be below $\mu_f (\ll P)$.
\pagebreak
Intuitively, short-distance physics at scale $P$ generates
fluctuations of color charges.
They are naturally regarded as 
superpositions of color multipoles
from the viewpoint of light quarks and gluons whose wave-lengths are
much larger than the fluctuation scale.
\begin{figure}
\begin{center}
\includegraphics[width=10cm]{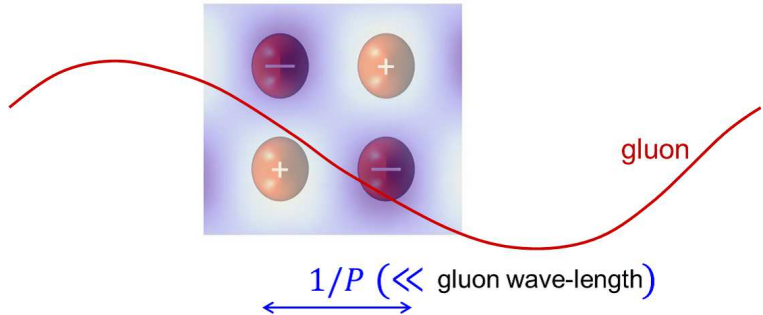}
\end{center}
\vspace*{-.5cm}
\caption{\small
Intuitive picture  of 
multipole expansion in an EFT.
Gluons in the EFT have wave-lengths much larger than
the typical scale $1/P$ of color-charge fluctuations.
Hence, it is natural to express
gluon fields by multipole expansion.
\label{MultipoleDistr}}
\end{figure}
See Fig.~\ref{MultipoleDistr}.
This is in analogy to classical electrodynamics, in which
an electric field at a large scale (compared to the
scale of charge distribution)
can be expressed generally as
a superposition of electric fields generated by 
electric multipoles.

In eq.~(\ref{OPE}) 
contributions from IR degrees of freedom
are contained in matrix elements of operators.
They cannot be evaluated by perturbative QCD
since dominant contributions come from the scale of order $\LQ$.
Rather these matrix elements are treated as non-perturbative
quantities
(parameters), which appear commonly in predictions of many observables.
Well-known examples are the local gluon condensate and chiral quark 
condensates, which are the lowest-dimension gauge-invariant
local operators of gluons and quarks, respectively.
These non-perturbative matrix elements are computed by lattice 
simulations or determined
phenomenologically from various experimental data.
Hence, IR renormalons in a purely perturbative prediction
of Sec.~\ref{Sec:PurelyPTQCD}
are replaced by (or absorbed into) the non-perturbative matrix elements
in the formulation of this section.
As a trade-off, the Wilson coefficients can be
predicted accurately (without IR renormalon uncertainties)
in this formulation.

\subsection{Dimensional regularization and IBP identities}
\clfn

Recent developments of perturbative QCD rely heavily
on  developments of various computational technologies.
We explain some features of dimensional regularization,
on which various modern technologies are based.

Let us discuss advantages and disadvantages of the
dimensional regularization.
Advantages can be stated as follows.
(1) This regularization preserves important symmetries
such as Lorentz symmetry and gauge symmetries.
(2) In a single step, we can render all loop integrals
and phase-space integrals finite, both at UV and IR.
This is in contrast to Pauli-Villars regularization, in which
one has to perform multiple subtractions according to
the degree of UV divergence of an original integral.
(3) There are many useful computational techniques which
are possible only with
this regularization \cite{Smirnov:2004ym}; we will see some examples below.

Disadvantages can be listed as follows.
(1) This regularization scheme is not well defined in the 
framework of quantum field theory.
Since we consider the space-time dimension $D$ as a
general complex variable, (at least naively) it is not 
possible to consider a quantum field theory in such a
dimension.
This is in contrast to e.g.\ lattice regularization, with
which a quantum field theory can be defined
unambiguously.
(Nevertheless, it may be worth notifying that
in perturbative computations the dimensional
regularization is well-defined and uniquely defined.)
(2) One faces many difficulties in physical interpretations.
There are typical questions beginners ask:
(a) Does $1/\epsilon^n$ represent IR or UV divergences?
Clearly poles in $\epsilon=(4-D)/2$ are not equivalent
to divergences, since one can write down integrals
which are UV or IR divergent at $D=4$
but free from poles in $\epsilon$.
Then, is it legitimate to identify poles in $\epsilon$ as UV
divergences in the usual procedure of renormalization?
(b) What is the meaning of seemingly unphysical equalities 
such as vanishing of a tadpole diagram
$\int d^Dk\, \frac{1}{k^2} =0$\,?
It is a UV divergent integral at $D=4$, but is there a rationale
or logical requirement
that it must be zero?

It is certainly helpful to be aware of these advantages
and disadvantages in trying to
find more precise interpretations of various predictions of perturbative QCD,
although it may not be necessary in carrying out practical
computations.\footnote{
It took years for the author to find interpretations to the
kinds of questions raised here.
Now he believes (with certain reasonings) that the
dimensional regularization leads to correct predictions
(which should not be dependent on the regularization you choose).
The reasoning helped in accomplishing tough computations.
}

Probably the most powerful application of dimensional regularization
is the integration-by-parts (IBP) identities \cite{Chetyrkin:1981qh}, which
we explain in the rest of this subsection.
The identities are derived from the Gauss theorem in $D$ dimension:
\bea
&&
\int d^Dp_1 \cdots d^Dp_L \,
\frac{\partial}{\partial X_\mu}
\left(
\frac{Y_\mu}{D_1^{n_1}\cdots D_N^{n_N}}
\right)
=0,
\\&&
X\in \{ p_1,\cdots,p_L\},
~~~
Y\in \{ p_1,\cdots,p_L, q_1,\cdots,q_M\}.
\eea
Here, $p_i$ and $q_i$ denote a loop momentum and an external
momentum, respectively;
$D_i$ denotes the denominator of an arbitrary propagator 
dependent on the internal and external momenta;
each power $n_i$ is an integer; in the case $n_i$ is negative,
$1/D_i^{n_i}$ represents a factor in the numerator
of the integrand.
The above equality is trivial in the case that the
dimension $D$ is a positive
integer and if the integral is finite, since
the surface term should vanish sufficiently quickly at infinity.
Conceptually it is an analytic continuation of
the equality to a complex dimension $D$ (although one needs to
do this carefully to be rigorous).
By operating the derivative before integration, one obtains
an identity among different types of loop integrals (see the
example below).
By choosing various $X$, $Y$ and $n_i$'s, one can generate
innumerable identities.
It is a standard technology of today's loop computations to use
these identities
to reduce a large number of loop integrals to a 
small set of simple integrals (master integrals).
For instance, it is not rare that order $10^3$--$10^4$ loop integrals
are reduced to order $10^1$--$10^2$ master integrals
by this reduction in contemporary loop computations.

\begin{figure}
\begin{center}
\vspace*{-.5cm}
\includegraphics[width=4.3cm]{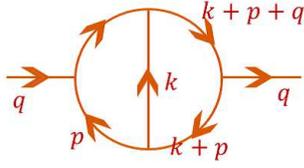}
\end{center}
\vspace*{-.5cm}
\caption{\small
Two-loop Feynman diagram of massless scalar particles,
which can be simplified using an IBP identity.
$q$ denotes the external momentum, while $k$ and $p$ denote the
loop momenta.
\label{2loopDiagram}}
\end{figure}
Let us take a diagram shown in Fig.~\ref{2loopDiagram} as an example and explain
the IBP identity.
All the internal lines are assumed to be massless.
One can generate a following identity:
\bea
&&
0=\int d^D\!p\,d^D\!k \, \frac{\partial}{\partial k^\mu}
\, \frac{k^\mu}{p^2k^2(k+p)^2(p+q)^2(k+p+q)^2}
\nonumber\\&&~
=\int d^D\!p\,d^D\!k \, 
\, \frac{1}{p^2k^2(k+p)^2(p+q)^2(k+p+q)^2}
\nonumber\\&&~
~~~~~~~~~~~~~~~~~~~~~
\times
\left[ 
D-\frac{2k\cdot k}{k^2}-\frac{2k\cdot(k+p)}{(k+p)^2}
-\frac{2k\cdot(k+p+q)}{(k+p+q)^2}
\right]
\nonumber\\&&~
=\int d^D\!p\,d^D\!k \, 
\, \frac{1}{p^2k^2(k+p)^2(p+q)^2(k+p+q)^2}
\left[ 
D-4+\frac{p^2-k^2}{(k+p)^2}
+\frac{(p+q)^2-k^2}{(k+p+q)^2}
\right] .
\nonumber\\&&~
\label{ExampleIBPid}
\eea
We suppressed $+i0$ in the propagator denominators.
In the last equality we expressed the numerators in
linear combinations of the propagator denominators.
\begin{figure}
\begin{center}
\includegraphics[width=16cm]{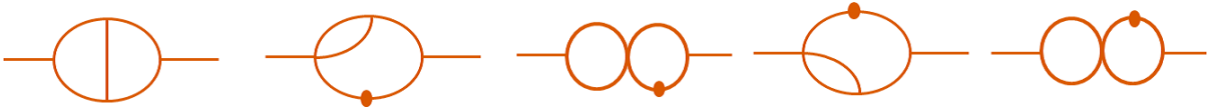}
\end{center}
\vspace*{-.5cm}
\caption{\small
Diagrams representing the IBP identity 
Eq.~(\ref{ExampleIBPid}).
The point on a line shows a two-point vertex, hence a line with
a point represents the square of the corresponding propagator.
\label{IBPid-2loopDiag}}
\vspace*{-.5cm}
\end{figure}
The last line can be regarded as a relation among
the diagrams shown in Fig.~\ref{IBPid-2loopDiag}.
Since some of the diagrams are the same, in fact the identity
relates three different diagrams.
Note that in the last line of eq.~(\ref{ExampleIBPid})
some of the propagator denominators are canceled by
the numerators.
As a result, the second and third diagrams of 
Fig.~\ref{IBPid-2loopDiag} have simpler topologies than
the first diagram.
Thus, the identity can be used to express the
first diagram in terms of two
diagrams with simpler topologies.

\subsection{Asymptotic expansion of diagrams and relation to EFT}
\clfn

In this subsection we explain the technique called asymptotic
expansion of a diagram or integration by regions \cite{Smirnov:2002pj}.
This can be used to identify operators ${\cal O}_i$
(effective interactions)
in the Lagrangian of a Wilsonian EFT, eq.~(\ref{EFT-Lagrangian}).
At the same time the technique provides an efficient method for
perturbative computations of Wilson coefficients $g_i(\mu_f)$.

Let us first explain the idea of the asymptotic expansion
in a simplified example.
We consider an integral
\bea
&&
I(m;\epsilon)=\int_0^\infty dp\, \frac{p^\epsilon}{(p+m)(p+1)} .
\label{defI}
\eea
It is a toy model imitating an
integral in dimensional regularization.
In fact, it is a one-parameter integral,
imitating integral over the radial direction of
a dimensionally-regulated integral, with
$p^\epsilon dp$ representing a volume element;
furthermore each propagator denominator is a linear
function of $p$ rather than a quadratic function.
Suppose $m \ll 1$ is a small parameter and consider
expanding $I$ in $m$.
Let us presume as if the integral region is
divided into two regions
$p< 1$ and ${p> 1}$, and
expand the integrand in each region in a small
parameter.
Nevertheless, we restore the original integral region in
each integral, as follows.
\bea
&&
I=
\int_0^\infty dp\, 
\underline{
\frac{p^\epsilon}{p+m}
(1-p+p^2+\cdots)
} ~+~
\int_0^\infty dp\, 
\underline{
\frac{p^\epsilon}{p+1}\,\frac{1}{p}
\left(1-\frac{m}{p}+\cdots\right) 
}.
\label{ExpansionOfI}
\\ &&
~~~~~~~~~~~
~~~~~~~~~~~
{p< 1}
~~~~~~~~~~~
~~~~~~~~~~~
~~~~~~~~~~~
~~~~~
{p> 1\gg m}
\nonumber
\eea
At a first glance, this seems to give a wrong result,
since firstly we have extended each integral region to a
region where the expansion is not justified, and secondly
there would be a problem of double counting of region.
Surprisingly, however, 
if we evaluate the individual terms of the above
integrals and take their sum, it gives 
the correct expansion in $m$ of the original integral $I$.

\begin{figure}
\begin{center}
\vspace*{-0.5cm}
\includegraphics[width=8.5cm]{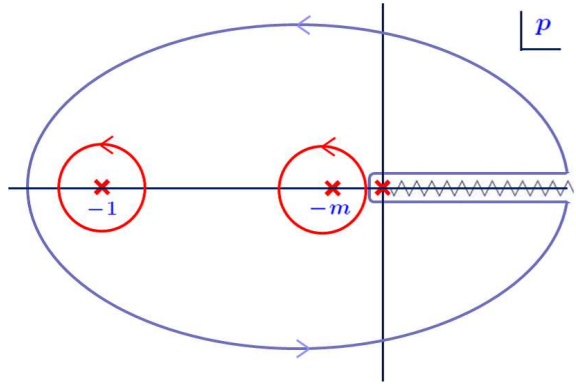}
\end{center}
\vspace*{-.5cm}
\caption{\small
Analyticity of
the integrand of eq.~(\ref{defI}) in the
complex $p$-plane.
The blue contour can be
deformed to the sum of the red contours.
\label{ContourAsymptExp}}
\end{figure}
The reason can be understood as follows.
Fig.~\ref{ContourAsymptExp} shows the analyticity of
the integrand of eq.~(\ref{defI}) in the
complex $p$-plane:
there are poles at $p=-1$ and $p=-m$;
the origin is a branch point due to
$p^\epsilon$ and the
branch cut lies along the positive $p$-axis.
The integral of $p$ along the positive $p$-axis
is equal to, up to a proportionality factor, 
an integral along the
contour wrapping the branch cut.
We may close the contour at negative infinity and deform the
contour into the sum of 
two closed contours surrounding the
two poles.
(See Fig.~\ref{ContourAsymptExp}:
we deform the blue contour
to the sum of the red contours.)
Along the contour surrounding the pole at $-m$, it is justified to
expand the integrand using the fact $|p|\approx m \ll 1$;
this gives the integrand of the first term of eq.~(\ref{ExpansionOfI}).
After the expansion, the contour of the integral of each term
of the expansion
can be brought back to the original contour wrapping the
branch cut along the positive $p$-axis.
Similarly, along the contour surrounding the pole at $-1$,
we may expand the integrand using $|p|\approx 1\gg m$,
which gives the second term of eq.~(\ref{ExpansionOfI}).
Again, after the expansion, the integral contour can be
brought back to the one surrounding the branch cut.\footnote{
In these manipulations, the value of $\epsilon$ in each term needs to
be varied appropriately by
analytical continuation into the domain where each integral
is well defined.
}
In this way, we obtain eq.~(\ref{ExpansionOfI}).

Thus, for an integral that imitates a dimensionally-regulated one,
we can expand the integral in a small parameter, 
without introducing a cut-off in the integral region.
The important point in the above example is that the contribution from
each of the scales $|p|\sim 1$ and $|p|\sim m$ is expressed
by a contour integral surrounding the corresponding pole
in the integrand (i.e., by the residue of each pole).

The method for the asymptotic expansion of a loop integral
in dimensional regularization is the same:
we divide the integral region into separate regions according to
the scales contained in the integrand and
expand the integrand in appropriate small parameters in
respective regions;
we nevertheless integrate individual terms of the expansions over
the original integral region, namely, over the entire $D$-dimensional
phase space for each loop integral.\footnote{
At the moment, 
the proof of this method using contour deformation 
as in the above toy model is missing,
for general loop integrals in dimensional regularization.
While it is likely that such an interpretation is possible generally,
presently this type of proof is valid only in some selective
cases.
There exists a general proof based on different reasonings
\cite{Jantzen:2011nz}.
}

\begin{figure}[h]
\vspace*{-.5cm}
\begin{center}
\includegraphics[width=14cm]{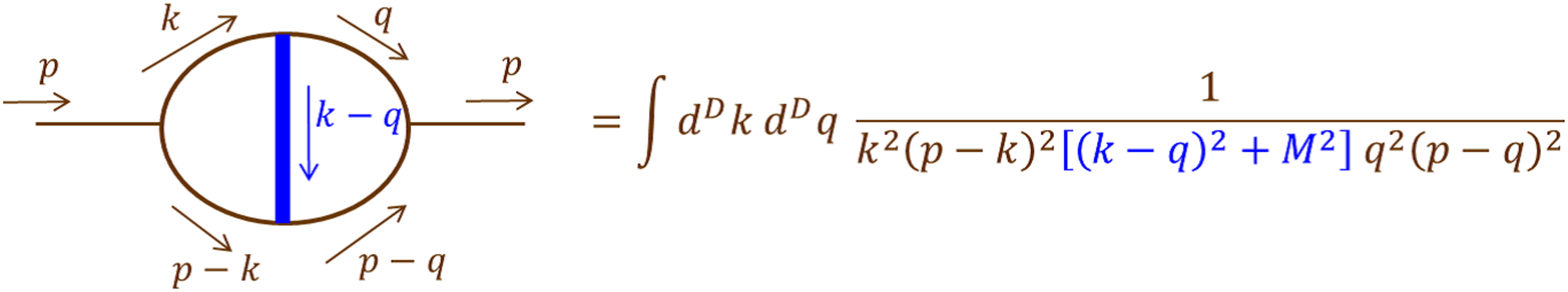}
\end{center}
\vspace*{-.5cm}
\caption{\small
Two-loop diagram used to illustrate asymptotic expansion
in $p^2/M^2$.
The thick blue line represents a propagator
with mass $M$, while all other lines represent massless
propagators.
\label{2loopDiag-AsymptExp}}
\vspace*{-.5cm}
\begin{center}
\includegraphics[width=16cm]{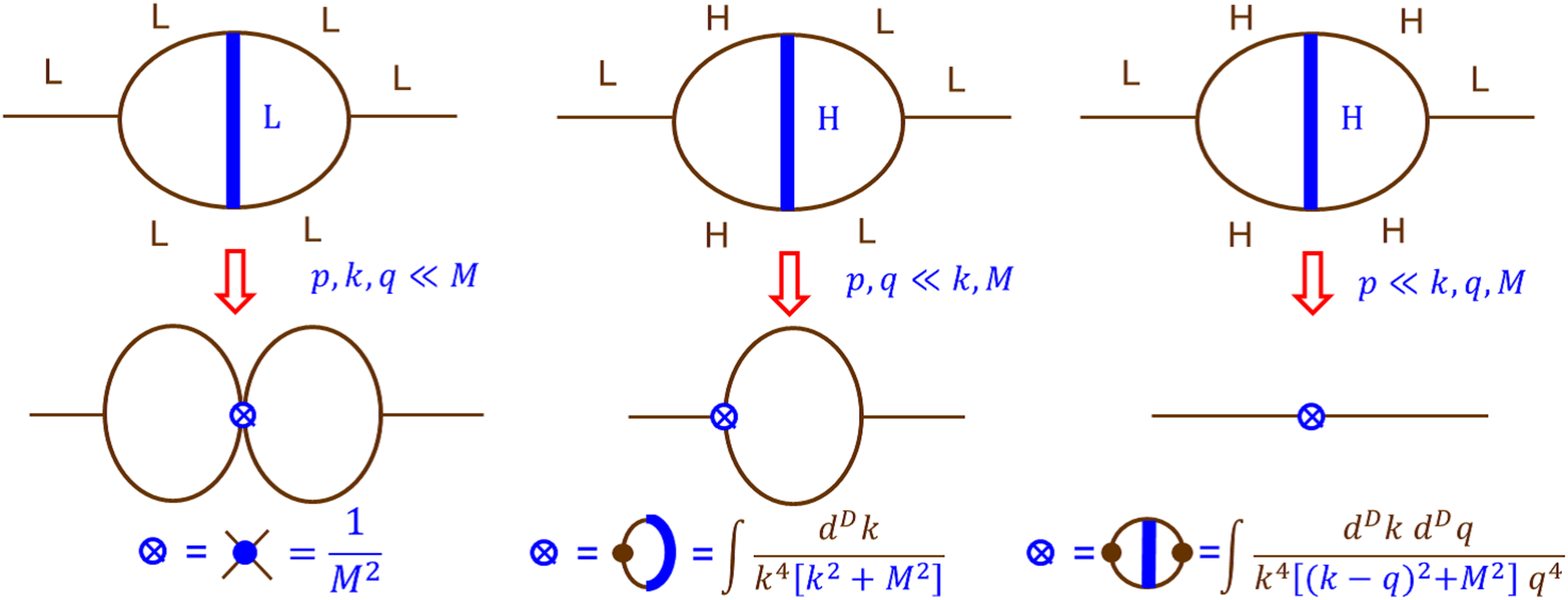}
\end{center}
\vspace*{-.5cm}
\caption{\small
Diagrams showing procedure of the asymptotic expansion.
The bottom line represent the Wilson coefficients
of the leading-order effective vertices in
respective regions.
\label{AsymptExpOf2loopDiag}}
\end{figure}

For illustration we consider the following two-loop integral
in the case $p^2\ll M^2$:
\bea
J(p^2,M^2)=
\int d^D\!k\,d^D\!q \, \frac{1}{k^2(p-k)^2[(k-q)^2+M^2]q^2(p-q)^2} .
\eea
The corresponding
diagram is shown in Fig.~\ref{2loopDiag-AsymptExp}, where the
thick blue line represents a heavy particle with
mass $M$ and all other lines
represent massless particles.
We expand $J$ in $p^2/M^2$.
The integral region of each loop integral is divided into two
regions: high momentum region (H), $|k|>M$ or $|q|>M$, and 
low momentum region (L), $|k|<M$ or $|q|<M$.
Hence, the whole integral region is divided into four regions:
(H,H),(H,L),(L,H),(L,L).
Of these (H,L) and (L,H) are the same due to the exchange symmetry
between $k$ and $q$.
Fig.~\ref{AsymptExpOf2loopDiag} shows how to perform the asymptotic
expansion in each of these regions.

In the region (L,L) we expand the massive propagator
$1/[(k-q)^2+M^2]$ in $k$ and $q$.
Each term represents an effective four-point vertex, where
the leading vertex is given by a constant coupling $1/M^2$.
This is depicted in the left-most part of the figure.
Higher-order vertices are associated with 
powers of the factor $(k-q)^2/M^2$, which correspond to
four-point interactions given by higher derivative operators.

\pagebreak
In the region (H,L) we expand the propagator $1/(p-k)^2$
in $p$ and the propagator $1/[(k-q)^2+M^2]$ in $q$.
In each term of the expansion, integral over $k$ can be factorized, 
since $p,q$ enter only the numerator of the integrand and can be
pulled outside of the integral.
This produces effective three-point vertices, which
correspond to three-point interactions given by
local operators.
The leading term of this expansion is depicted in the
middle part of the figure.
Since high momenta flow through the $k$-loop,
it is natural to expect that the loop 
effectively shrinks to a point.

In the region (H,H) we expand $1/(p-k)^2$ and $1/(p-q)^2$
in $p$.
In this case, the whole integral over $k$ and $p$ can be
factorized at each order of the expansion.
Thus, each term can be regarded as an effective two-point
interaction corresponding to a local operator.
See the right-most part of the figure.

We may compute the same process in a low-energy EFT in which the massive
particle has been integrated out.
The asymptotic expansion of the diagram in the full theory obtained above
can be interpreted as the computation in the EFT.
The bottom-left diagram in Fig.~\ref{AsymptExpOf2loopDiag} represents
a two-loop computation of this process in the EFT with an insertion of
a four-point vertex, which is generated at tree-level of the full
theory.
The factor $1/M^2$ below the diagram represents the Wilson coefficient
of the leading-order vertex in expansion in $1/M^2$.
The bottom-middle diagram represents a one-loop computation of this
process in the EFT
with an insertion of a three-point vertex, which is generated at
one-loop level in the full theory.
The one-loop integral shown below the diagram represents the
Wilson coefficient of the leading-order vertex in expansion in $1/M^2$.
The bottom-right diagram represents a tree-level computation of this
process in the EFT with an insertion of a two-point vertex, which
is generated at two-loop level in the full theory.
The corresponding leading-order Wilson coefficient is shown
as a two-loop integral.
Thus, the relevant operators and Wilson coefficients of EFT can be
identified.

The Wilson coefficients, given by loop integrals in 
dimensional regularization, are 
particularly convenient in practical computations.
They are homogeneous in a single dimensionful parameter $M$, 
which can be computed relatively easily.
In contrast, if we adopt a cut-off regularization,
usually it becomes much more difficult to evaluate the 
corresponding integrals
(especially at higher loops),
since more scales are involved.

\subsection{Higher-loop integrals and multiple zeta values}
\clfn

In this subsection we explain developments and recent
activities in  analytic evaluations of
higher-loop corrections.
As already mentioned, a general algorithm for
analytic evaluation has not been found
yet and there are diverse on-going researches.
In particular we focus on intersection with an area of research 
in mathematics on multiple zeta values (MZVs).

We start by looking at 
analytic expressions for multi-loop corrections to  
the anomalous magnetic moment of electron
($g_e-2$) as an example:\footnote{
Although it is a perturbative QED prediction, its
analytic feature, in which we are interested, is similar
to that of perturbative QCD predictions.
}
\bea
&&
a_e({\rm QED})= a_e^{(2)}\Bigl(\frac{\alpha}{\pi}\Bigr)+
a_e^{(4)}\Bigl(\frac{\alpha}{\pi}\Bigr)^2+
a_e^{(6)}\Bigl(\frac{\alpha}{\pi}\Bigr)^3+
a_e^{(8)}\Bigl(\frac{\alpha}{\pi}\Bigr)^4+
a_e^{(10)}\Bigl(\frac{\alpha}{\pi}\Bigr)^5+
\cdots ,
\\ &&
a_e^{(2)}=\frac{1}{2} ,
\\ &&
a_e^{(4)}=\frac{197}{144}+\frac{\pi^2}{12}+\frac{3}{4}\zeta(3)
-\frac{\pi^2}{2}\ln 2 ,
\\ &&
a_e^{(6)}=\frac{83}{72}\pi^2\zeta(3)-\frac{215}{24}\zeta(5)+\frac{100}{3}
\left({\rm Li}_4\Bigl(\frac{1}{2}\Bigr)+\frac{1}{24}\ln^4 2 -\frac{1}{24}\pi^2\ln^2 2
\right)
-\frac{239}{2160}\pi^4
\nonumber\\&&~~~~~~~~
+\frac{139}{18}\zeta(3)-\frac{298}{9}\pi^2\ln 2+\frac{17101}{810}\pi^2+\frac{28259}{5184},
\eea
where ${\cal O}(m_e/m_\mu)$ terms are omitted.
Presently the corrections are known analytically 
up to three loops ($a_e^{(6)}$) \cite{Laporta:1996mq}
while they are known numerically up to five loops ($a_e^{(10)}$)
\cite{Aoyama:2012wj}.
The above expressions include a
class of transcendental numbers, which can be expressed as
infinite sums:
\bea
&&
\zeta(n)=\sum_{m=1}^\infty \frac{1}{m^n},
~~~~~
\ln 2=-\sum_{m=1}^\infty \frac{(-1)^m}{m},
~~~~~
\\&&
-2\left({\rm Li}_4\Bigl(\frac{1}{2}\Bigr)+\frac{1}{24}\ln^4 2 -\frac{1}{24}\pi^2\ln^2 2
\right)+\frac{\pi^4}{180}=
\sum_{m>n>0} \frac{(-1)^{m+n}}{m^3 n}, 
\eea
where $\zeta(x)$ and 
${\rm Li}_n(x)=\sum_{k=1}^\infty \frac{x^k}{k^n}$ denote 
the Riemann zeta function and polylogarithm, respectively.
Note that, since $\zeta(2n)$ is proportional to $\pi^{2n}$ for 
$n\in \mathbb{N}$, $\pi^2$ and $\pi^4$ in 
$a_e^{(2n)}$ can be regarded as this class of numbers
(up to rational coefficients) as well.

A generalized MZV is defined by a nested sum
\bea
Z\left(\infty;a_{1},a_{2},\ldots,a_{N};\lambda_{1},\lambda_{2},\ldots,\lambda_{N}\right) =
\!\!\!\!\!
\sum_{ n_{1}>n_{2}>\cdots>n_{N}>0}\frac{\lambda_{1}^{n_{1}}\lambda_{2}^{n_{2}}\cdots\lambda_{N}^{n_{N}}}{n_{1}^{a_{1}}n_{2}^{a_{2}}\cdots n_{N}^{a_{N}}},
\label{generalizedMZV}
\eea
where
$a_i\in \mathbb{N}$ and $a_1\ge 2$.
In the case $\lambda_i \in \{1\}$ it is simply called a MZV;
in the case $\lambda_i \in \{-1,1\}$ it is called a sign-alternating
Euler sum.
We also consider the
case that $\lambda_i$ is a root of unity (e.g.\ $\lambda_i=e^{i\pi/3}$).
In many classes of higher-loop computations, including
the above example,
these types of generalized MZVs
appear in the analytic results.

Each generalized MZV also has a nested integral representation.
By way of example,
\bea
\int_0^1 \frac{dx}{x}\int_0^x\frac{dy}{y-\alpha}\int_0^y\frac{dz}{z-\beta}
=-Z\Bigl(\infty;2,1;\frac{1}{\alpha},\frac{\alpha}{\beta}\Bigr) ,
\eea
which can be verified easily by rewriting the integrand
as an infinite series expansion in $y$ and $z$ and integrating
each term.
As inferred from this example, 
in the case that $\lambda_i$'s are roots of unity, generally
the nested integral representation
has singularities of
the integral variables at zero or at roots of unity.
We will use this property below.

Hereafter, we refer to generalized MZV simply as MZV.
The weight of a MZV is defined as the sum of the powers of the summation
indices in the denominator: $w=a_1+\cdots + a_N$ for eq.~(\ref{generalizedMZV}).
It turns out that various MZVs can be expressed by a small set of
basis MZVs.
More precisely, one may consider 
the vector space over $\mathbb{Q}$ spanned by
MZVs at each given weight, and the dimension of the vector
space is much smaller than the number of MZVs.
For instance,
since the relation 
$\displaystyle
\sum_{m>n>0} \frac{1}{m^2n}=\sum_{m=1}^\infty \frac{1}{m^3}=\zeta(3)
$
holds,
we find that, for $\lambda_i\in\{1\}$ and at weight three, 
the dimension is one, while there
are two MZVs.
In fact, in the case $\lambda_i\in\{1\}$, mathematicians
have proven \cite{Thm1,Thm2,Thm3} that the dimension of the vector space at
weight $w$ is less than or equal to $d_w$, which is 
determined by a Fibonacci-type recursion relation
\bea
d_0=1,~~~d_1=0,~~~d_2=1,~~~d_w=d_{w-2}+d_{w-3}~(w\ge 3) .
\eea
\pagebreak
(The dimension is most likely to be equal to $d_w$.)
Tab.~\ref{tab:dimMZV} shows $d_w$ and the number of MZVs for weights
up to 12.
One sees that indeed MZVs can be expressed by a small set of basis.
This fact allows us to obtain compact analytic expressions
for higher-loop radiative corrections, although they originally
stem from vast numbers of Feynman diagrams.
\begin{table}
\begin{center}
\small
\begin{tabular}{|c|l|r|r|r|r|r|r|r|r|r|r|r|r|r|r|r|r|}
\hline
weight&$w$&0&1&2&3&4&5&6&7&8&9&10&11&12\\
\hline
dimension(?)&$d_w$&1&0&1&1&1&2&2&3&4&5&7&9&12\\
\hline
\#(MZVs)&$2^{w-2}$&--&--&1&2&4&8&16&32&64&128&256&512&1024\\
\hline
\end{tabular}
\end{center}
\caption{\small
(Upper bound for) Dimension $d_w$ of the vector space
spanned by MZVs and the
number of MZVs for a given weight $w$ up to 12
and $\lambda_i\in \{1\}$.
\label{tab:dimMZV}}
\end{table}
Properties of MZVs can be explored using various 
non-trivial algebraic
relations among them.
In particular, empirically ``shuffle relations'' are 
known to be powerful
in reducing MZVs to a smaller set
(and probably provide maximal algebraic relations
in the case $\lambda_i\in\{1\}$)
\cite{ShuffleRel}.\footnote{
We have recently found a new type of relations independent
of the shuffle relations in the case $\lambda_i$'s are
roots of unity \cite{Anzai:2012xw}.
}
It is stated that, by interpreting
MZVs as periods of
mixed motives, properties
of MZVs can be understood from deeper viewpoints.
Indeed mathematicians seem to observe deeper structures
behind the world of MZVs.

Relations between topologies of Feynman diagrams and MZVs are
the subjects of common interests for mathematicians and
physicists; see e.g.\ \cite{Broadhurst:1996az,Broadhurst:1998rz,Brown:2010bw}.
Mathematicians may want to relate a topology, represented by
a Feynman diagram, to MZVs as given by the value
of the diagram.
This may be used to characterize topologies if the relations
are understood well.
Physicists want to find a systematic way to
relate a topology of a diagram to the value of the diagram
when expressed in terms of MZVs.
If this is achieved, we may find an efficient way to
evaluate Feynman diagrams in terms of MZVs.

\begin{figure}
\begin{center}
\includegraphics[width=6cm]{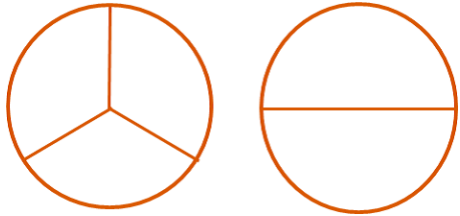}
\end{center}
\vspace*{-.5cm}
\caption{\small
Diagrams which include MZVs with sixth roots of unity
$\lambda_i=e^{in_i\pi/3}$.
Each line corresponds to either a massive propagator with mass $m$ or 
a massless propagator.
\label{BenzDiagram}}
\end{figure}

Below we discuss what kind of MZVs (in particular 
which $\lambda_i$'s) are associated with a diagram.
Although no systematic argument exists up to now,
we discuss some empirical aspects gained through experiences
in explicit higher-loop computations.
For instance, the diagrams shown in Fig.~\ref{BenzDiagram}
are known to be expressed by MZVs 
with $\lambda_i$'s given by sixth roots of unity.

In general,
$\lambda_i$'s are closely connected to singularities in a Feynman
diagram, which are also closely tied to the topology
of the diagram.
Singularities contained in a Feynman diagram may be
classified as follows:
\begin{itemize}
\item
IR singularity, generated as external momenta are taken to zero.
\item
UV singularity, generated as external momenta are taken to infinity.
\item
Mass singularity, generated as the masses
of internal particles are taken to zero.
\item
Threshold singularity, generated as 
the external energy crosses the
threshold energy of an intermediate state.
\end{itemize}
In general 
overlaps of these types of singularities are included
in a multiloop digram.
There is no known systematic method to disentangle
general overlapping singularities in such a
way to cast them into MZVs (if this is possible at all).\footnote{
In numerical evaluations
disentanglement of overlapping singularities 
is always possible in finite steps, using 
sector decomposition and the theorem on resolution of singularities
\cite{Bogner:2007cr}.
}
In simple cases this can be achieved, for instance,
by iteratively applying the method of differential equation 
\cite{Remiddi:1997ny};
one can reduce a complicated diagram as integrals over
simpler diagrams iteratively, and in the end as a combination of
nested integrals, which can be converted to MZVs.

\begin{figure}
\begin{center}
\begin{tabular}{cc}
\includegraphics[width=5cm]{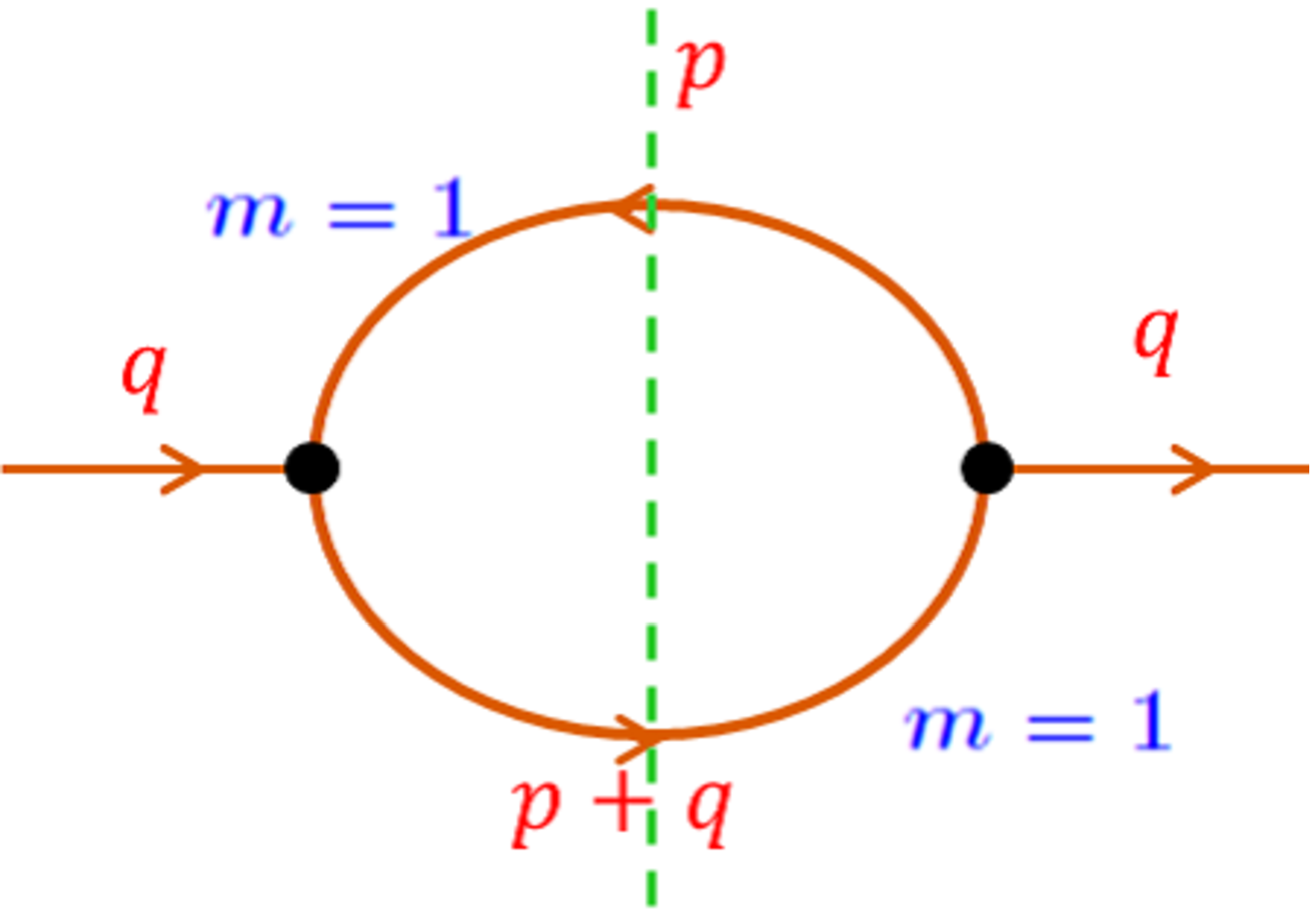}
&
\includegraphics[width=8cm]{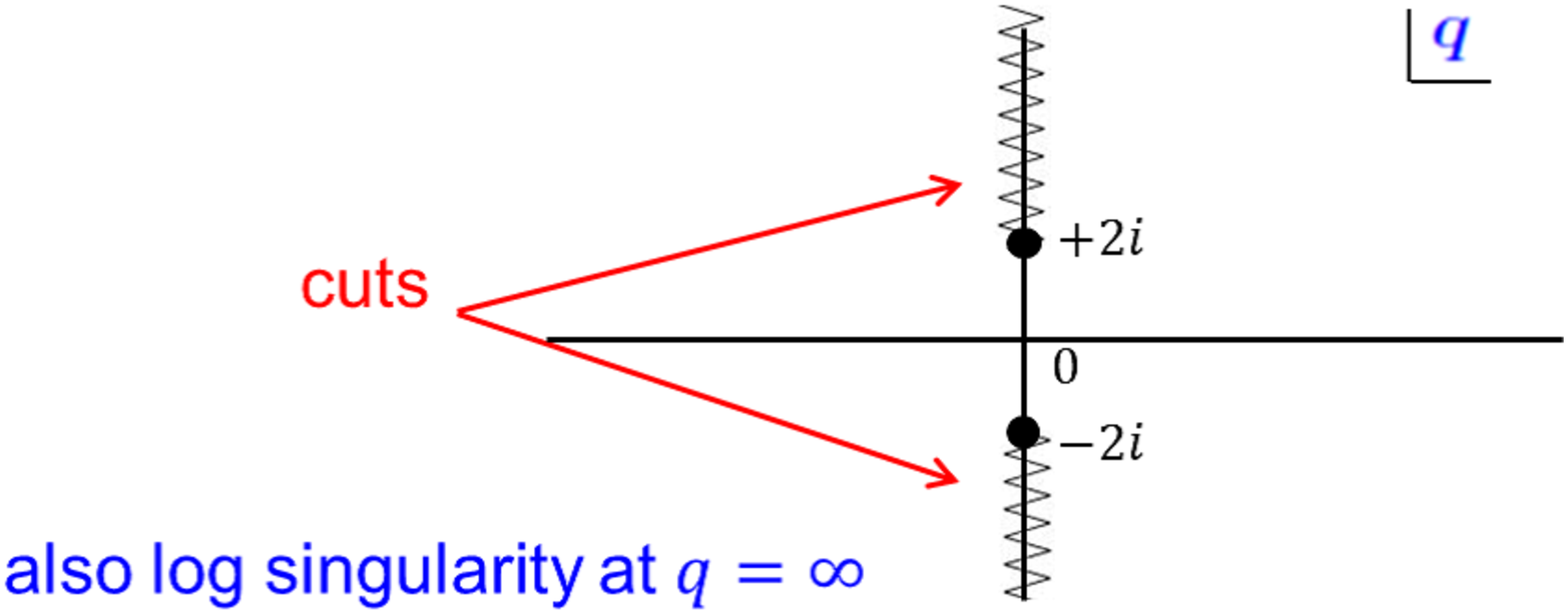}\\
(a)&~~~~~~~~~~~~~(b)
\end{tabular}
\end{center}
\vspace*{-.5cm}
\caption{\small
(a) Diagram for $I(q)$ defined in Eq.~(\ref{1loop2ptfn}).
The mass in the propagators is taken as $m=1$.
Imaginary part of $I(q)$ appears for $q^2<-(2m)^2=-4$ corresponding
to the cut shown by the dashed line.
(b) Analyticity of
$I(q)$ in the complex $q$-plane.
\label{1LoopVacPol}}
\end{figure}
For illustration we consider a simple example.
Consider a one-loop integral
\bea
I(q)=\int d^4p\, \frac{1}{(p^2+1)^2[(p+q)^2+1]}
\propto \frac{1}{q^2\sqrt{1+4/q^2}}\,\log
\left(\frac{\sqrt{1+4/q^2}+1}{\sqrt{1+4/q^2}-1}\right) .
\label{1loop2ptfn}
\eea
The diagram is shown in Fig.~\ref{1LoopVacPol}(a), where
the masses of internal lines are taken to be the same,
$m=1$.
The power of one of the propagators is raised in order to make
the integral well defined in four dimension.\footnote{
Powers of propagators are not essential in the following argument
to identify $\lambda_i$ of MZV.
}
Fig.~\ref{1LoopVacPol}(b) shows the analyticity of
$I(q)$ in the complex $q$-plane.
The square-root $\sqrt{1+4/q^2}$ generates branch points
corresponding to the threshold singularities at $q=\pm 2im=\pm 2i$.
(Note that we are working in Euclidean space-time.)
There is also a logarithmic singularity at $q=\infty$, which
can be regarded as a UV singularity, or alternatively 
as a mass singularity.

We can express the second diagram of Fig.~\ref{BenzDiagram}
as an integral over $I(q)$:
\bea
\int d^4 q\, \frac{1}{(q^2+1)^2}\,I(q) \propto
{\rm Im}\left[
\sum_{n=1}^\infty \frac{e^{i\pi n/3}}{n^2}
\right] ,
\eea
where we raised the power of the propagator to
make the integral well defined.
All the lines of the diagram have equal masses, $m=1$.
On the left-hand-side,
we can convert the integral to a one-dimensional integral over $q$.
The square-root in $I(q)$ can be eliminated by an Euler transformation
$x=\frac{1}{2}(1+\sqrt{1+4/q^2})$.
We may further express the logarithm as an integral
of a rational function.
In this way, one can convert the integral to a
nested double integral of rational functions, which is
in fact a MZV.
The map of the singularities by the Euler transformation 
is given by
\bea
\{\pm i ,\pm 2i,\infty \} 
\xrightarrow[~\mathlarger{x=\frac{1+\sqrt{1+4/q^2}}{2}\rule{0mm}{7mm}}~~~]{}
\{ e^{\pm\pi i/3},\mbox{---},(0,1) \}.
\eea
The singularities at $\pm i$ are associated with the
propagator $1/(q^2+1)^2$ attached to $I(q)$ and those
at $\{\pm 2i,\infty\}$ are the ones associated with $I(q)$.
The map is double-valued, and the singularities at
$\pm 2i$ are mapped to regular points.
Thus, we see that sixth-roots of unity are generated
by this map from physical singularities in the
diagram.

Empirically we observe similar mechanisms in relating
physical singularities of a diagram to $\lambda_i$'s of MZVs, 
for instance, when
conversion to MZVs is possible  by
the method of differential equation.
Sometimes square-roots are involved at intermediate
stages, and
in simple cases all the square-roots can be 
eliminated by successive Euler transformations.
Thus, the origins of $\lambda_i$'s are indeed attributed
to physical singularities
of diagrams in these cases.

\subsection{Closing the first part: summary and benchmarks}

\begin{figure}[h]
\begin{center}
\includegraphics[width=14cm]{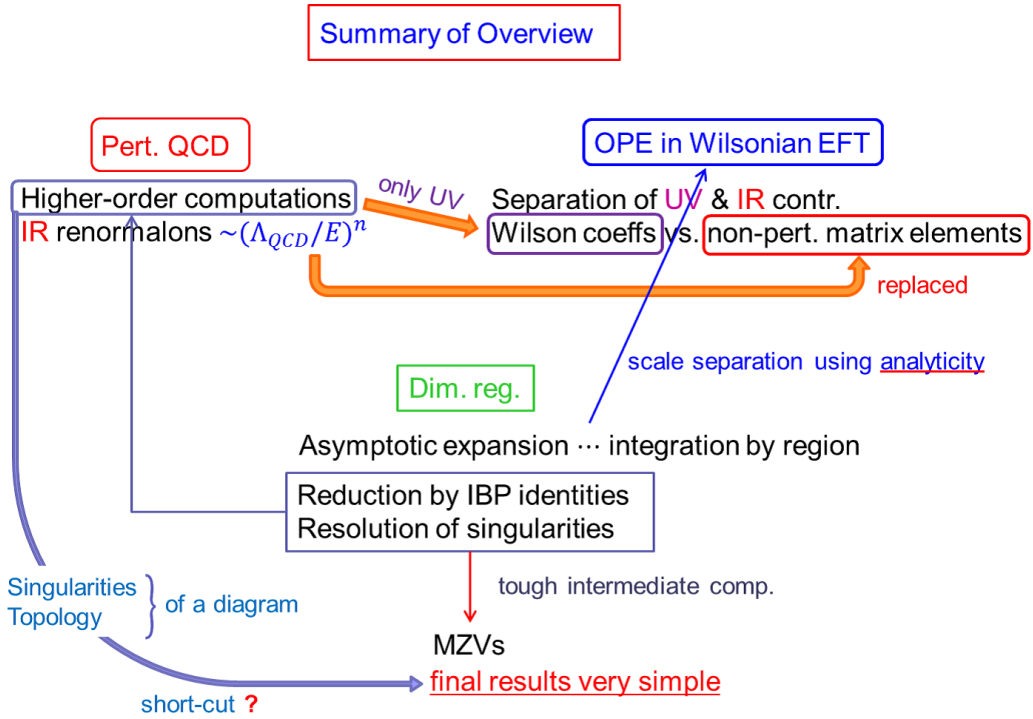}
\end{center}
\vspace*{-.5cm}
\caption{\small
Schematic diagram showing mutual connections 
of the subjects covered in the overview
given in Secs.~2 and 3.
\label{Summary1stPart}}
\end{figure}

Let us summarize the overview of perturbative QCD 
presented above.
This is depicted schematically in Fig.~\ref{Summary1stPart}.
Higher-order corrections in purely perturbative predictions
include uncertainties of order
$(\LQ/E)^n$ from IR renormalons.
This is because higher-order
corrections tend to be more sensitive to IR regions, 
which can effectively be represented by
an increase of the strong coupling
constant $\alpha_s(k)$ at IR.
Using OPE within a Wilsonian EFT, one may separate UV and IR
contributions.
The former are incorporated into
Wilson coefficients and the latter into
non-perturbative matrix elements.
In this formulation, only UV part of the higher-order
corrections of the purely perturbative prediction
is encoded in the Wilson coefficients, whereas 
uncertainties originating from IR renormalons are replaced
by non-perturbative matrix elements.
Hence, intrinsic uncertainties of perturbative expansions can be
eliminated,
and we obtain more accurate predictions as we compute
higher-orders corrections (as far as non-perturbative matrix elements
can be determined in some way).
These frameworks have been known since long time, at least
conceptually.
In practice, however, higher-order computations
based on EFTs had been quite difficult, since in
earlier days they were formulated by introducing a cut-off 
in the integrals in an essential way.

Developments of perturbative QCD over the last few decades rely
heavily on significant developments in computational technologies.
In particular, dimensional regularization has been the theoretical basis
for many important technological developments.

Asymptotic expansion of diagrams, or integration by regions, 
is a technique based on dimensional regularization and has
provided a solid foundation to OPE of EFT.
This technique enables computations of Wilson coefficients to
high orders, since integrals can be cast into forms much easier
to evaluate than ones with a cut-off.
An important feature is that the scale separation is realized
using analyticity of diagrams.
A contour integral surrounding each
singularity represents a contribution from the corresponding scale.

Furthermore,
we explained a reduction of loop integrals
using IBP identities,
and discussed relation between singularities of Feynman diagrams
and MZVs.
These techniques are used in higher-loop computations.
Essentially the computations are composed by
processes of resolution of singularities
in Feynman diagrams.
As a general feature of these processes,
the amount of calculation expands enormously at intermediate
stages.
On the other hand, final results tend to be compact, 
reflecting the nature of MZVs.
This indicates that there may be a short-cut in finding
final results, for instance by looking at singularities
and topologies of diagrams.

Thus, dimensional regularization
has brought in new computational methods 
as well as viewpoints which are complementary and contrasting 
to those based on cut-off regularization.
Fig.~\ref{Summary1stPart} displays a summary of
the current status of perturbative QCD as seen from the
author's viewpoint.\footnote{
On-going active developments related to
LHC physics are omitted.
}
This sets up a general formulation of analyses.
We should discuss physics using the formulation.
As an example, we analyze heavy quarkonium systems 
in the second part of this paper.
\medbreak

To end the first part, it would be fruitful to
note some of today's benchmarks of
perturbative QCD.
\vspace*{2mm}
\\
\underline{Universality}

\begin{figure}
\begin{center}
\includegraphics[width=7.5cm]{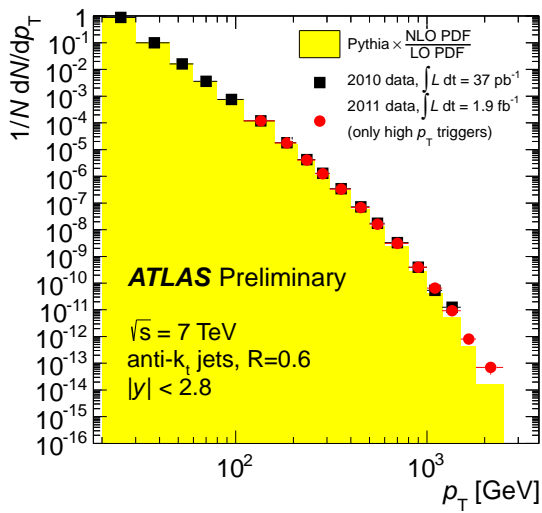}
\end{center}
\vspace*{-.5cm}
\caption{\small
Jet $P_T$ spectrum measured 
at the LHC experiment at $\sqrt{s}=7$~TeV \cite{ATLASdata}.
Observed inclusive jet $P_T$ distribution (black dots) is  
compared to MC prediction (yellow histogram). 
The distribution is normalized to unity and only 
statistical uncertainties are included.
\label{ATLASjetPTspect}}
\end{figure}
In the limit where the quark masses are neglected, QCD is a
theory with a single scale $\LQ$.
It is remarkable that such a theory can explain
physics phenomena over many orders of magnitude.
Fig.~\ref{ATLASjetPTspect} shows a comparion of 
experimental data and perturbative QCD prediction 
(in the sense of type (c) in Sec.~2.1) for the
jet $P_T$ spectrum at LHC.
The theory prediction is predominantly determined by
perturbative QCD.
The current theoretical errors
are typically order 10\% to a few tens \%.
In the figure jet $P_T$ ranges over two orders of magnitude
(from 10~GeV to a few TeV), while the
inclusive jet $P_T$ distribution varies more than
ten orders of magnitude!
Although the relative accuracy of the prediction at each
$P_T$ is not so precise, it is quite impressive
to observe a good agreement between the perturbative QCD
prediction and experimental
data for such a wide range of variation.
\medbreak

\noindent
\underline{Precisions}

Presently
several parameters of the QCD Lagrangian
are determined 
using perturbative QCD, combined with
experimental data and results of lattice QCD computations,
with the following accuracies \cite{Beringer:1900zz}:
\bea
&&
\alfs(M_Z)=0.1184(7)  \,~~~~~~~~ 0.6\%~{\rm accuracy},
\\ &&
\overline{m}_b=4.18(3)~~{\rm GeV} ~~~~ ~~~~~\, 0.8\%~{\rm accuracy},
\\ &&
\overline{m}_c=1.275(25)~{\rm GeV}  ~~ ~~~~~\, 2\%~~~{\rm accuracy},
\\ &&
\overline{m}_t=160^{+5}_{-4}~~{\rm GeV} ~~~~~~ ~~~~~\, 3\%~{\rm accuracy}
~~(\to 0.06\% ~~\mbox{at ILC}),
\eea
where $\alfs(M_Z)$ denotes the strong coupling constant
in the modified minimal-subtraction ({\msbar}) scheme, renormalized at the
scale of the mass of the $Z$ boson;
$\overline{m}_q$ denotes the {\msbar} mass 
(mass defined in the {\msbar} scheme)
renormalized
at the {\msbar} mass scale of a quark $q$.\footnote{
The current top quark mass $m_t=173.34\pm 0.76$~GeV \cite{ATLAS:2014wva}
measured at Tevatron and LHC is theoretically
not well defined.
The measured mass corresponds to a parameter in MC
simulations and its relation to the
parameters of the QCD Lagrangian is unknown beyond order 1--2~GeV accuracy.
By contrast, the {\msbar} mass is well defined and its relation
to the parameters of the QCD Lagrangian is theoretically solid. 
}
As can be seen, perturbative QCD is entering an era of
high precision science.
\vspace*{0.7cm}

\section{Application to Heavy Quarkonium Systems}

In this section we review application of perturbative
QCD to heavy quarkonium systems with
respect to the formulation discussed 
in Secs.~2 and 3.
In particular we discuss
(1) ${\cal O}(\LQ)$ physics in the heavy quark mass and
interquark force, and 
(2) renormalization of Wilson coefficients at IR.

\subsection{IR contributions in the static QCD potential}
\clfn

\begin{figure}[h]
\begin{center}
\includegraphics[width=12cm]{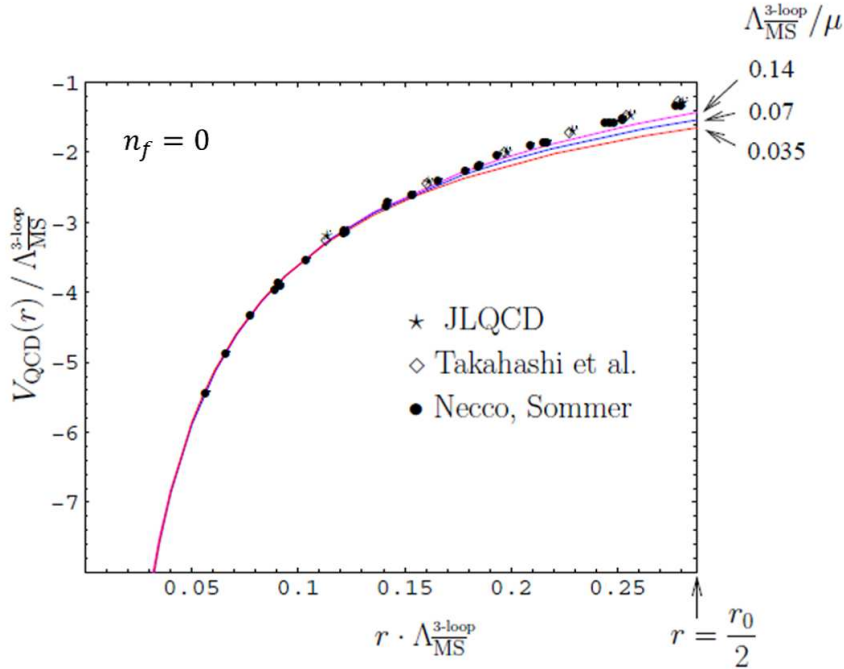}
\end{center}
\vspace*{-.5cm}
\caption{\small
Static QCD potential as a function of the distance between the static
charges $r$.
Both axes are scaled by powers of
$\Lambda_{\overline{\rm MS}}^{\mbox{\scriptsize 3-loop}}$.
Solid lines represent NNNLO perturbative QCD predictions
with different scale choices.
Data points represent lattice computations by three
different groups.
\label{3LoopQCDPot}}
\vspace*{0.5cm}
\end{figure}

Let us first quote the current status of the static QCD potential.
Fig.~\ref{3LoopQCDPot} shows the potential
energy between two static color charges, which combine to
color-singlet, as a function of
the distance
$r$ between the charges.
The units are scaled by the QCD scale in the
{\msbar} scheme at
three-loop order, 
$\Lambda_{\overline{\rm MS}}^{\mbox{\scriptsize 3-loop}}$
\cite{Chetyrkin:1997sg}.
The
NNNLO perturbative QCD prediction and lattice computations are
compared.
The three solid lines correspond to the perturbative predictions
\cite{Anzai:2009tm,Smirnov:2009fh}
with scale choices\footnote{
It is customary to vary the renormalization scale $\mu$ by
a factor 2 or 1/2 in estimating uncertainties of 
perturbative QCD predictions.
}
$\Lambda_{\overline{\rm MS}}^{\mbox{\scriptsize 3-loop}}/\mu
= 0.14$, 0.07 and 0.035.
The data points represent lattice results by three
different groups 
\cite{Necco:2001xg,Takahashi:2002bw,Aoki:2002uc}.
The number of quark flavor is set to zero in both computations.
$r_0$ denotes the Sommer scale, which is interpreted as about 0.5~fm.
\pagebreak
Hence, the largest $r$ in this figure is about 0.25~fm
[$\approx (0.8~{\rm GeV})^{-1}$].
Since the relation between the lattice scale ($r_0$) and
$\Lambda_{\overline{\rm MS}}^{\mbox{\scriptsize 3-loop}}$ 
is taken from other source,
the only adjustable parameter in this comparison is
an $r$--independent constant added to each potential,
whose value is chosen such that
all the potentials coincide at
$r \Lambda_{\overline{\rm MS}}^{\mbox{\scriptsize 3-loop}} = 0.1$.
We see a good agreement between
the perturbative and lattice predictions in the displayed
range.

To understand the nature of the perturbative prediction for 
the static potential $V_{\rm QCD}(r)$,
we first discuss the IR renormalons of $V_{\rm QCD}(r)$.
Since the perturbative prediction is more accurate at short distances,
$r\ll\LQ^{-1}$,
we consider (naively) a short-distance expansion of
$V_{\rm QCD}(r)$:
\bea
V_{\rm QCD}(r)\sim\frac{c_{-1}}{r} + c_0 + c_1r + c_2 r^2 +\cdots .
\label{naive-expansion}
\eea
This
expansion in $r$ is naive, in the sense that there must be a logarithmic
correction (at least) to the Coulomb term
$\sim 1/[r \log (\LQ r)]$, as designated by the
renormalization-group equation.
\begin{figure}
\vspace*{.5cm}
\begin{center}
\begin{tabular}{cc}
\includegraphics[width=5cm]{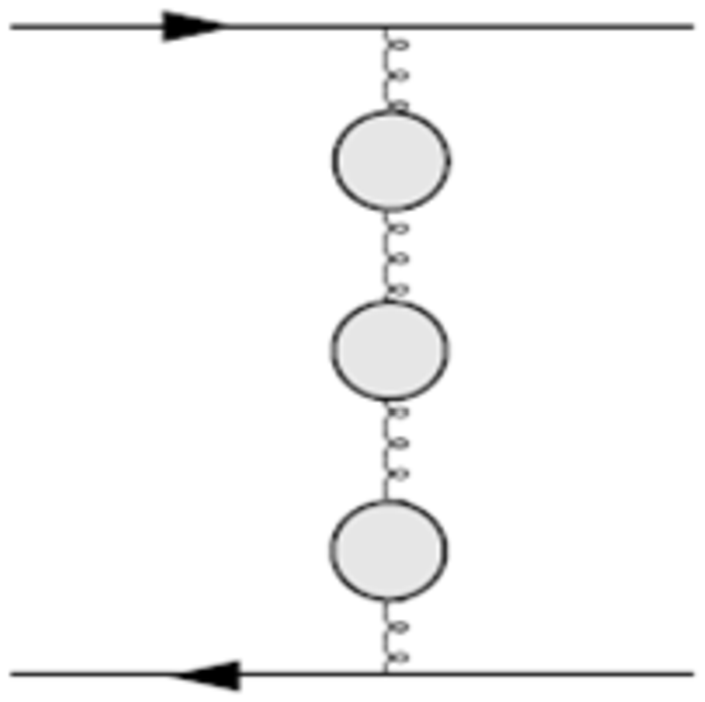}
&
\includegraphics[width=5.5cm]{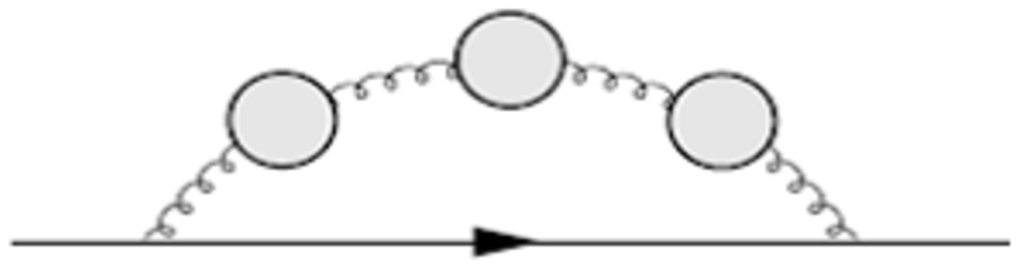}\\
(a)~~~~~&(b)
\end{tabular}
\end{center}
\caption{\small
Bubble-chain diagrams contributing to (a) the static QCD potential
and (b) the pole mass.
\label{Pot-SE-BubbleChain}}
\end{figure}
\pagebreak
From the computation 
of the bubble-chain diagrams in Fig.~\ref{Pot-SE-BubbleChain}(a)
IR renormalons in the perturbative
prediction for $V_{\rm QCD}(r)$ can be estimated
\cite{Aglietti:1995tg}; see Sec.~2.1.
They arise in the form $\LQ (\LQ\, r)^P$.
The leading uncertainty at $r\ll\LQ^{-1}$
is included in the $r$-independent constant, while the next-to-leading
uncertainty is included in the $r^2$ term:
\bea
&&
c_0 ~~~\sim {\cal O}(\LQ) ,
\\ &&
c_2\, r^2 \sim {\cal O}(\LQ^3r^2) .
\eea

The leading uncertainty can be eliminated in the following manner.
Consider the total energy of a static quark pair defined by
\bea
E_{\rm tot}(r)=2m_{\rm pole}+V_{\rm QCD}(r) .
\eea
We express the quark pole mass $m_{\rm pole}$ by the
{\msbar} mass $\overline{m}\equiv
m_{\overline{\rm MS}}(m_{\overline{\rm MS}})$ as
\bea
m_{\rm pole}=\overline{m}\,(1+d_1\alfs + d_2\alfs^2+d_3\alfs^3
+d_4\alfs^4+\cdots ) .
\label{PoleMSbarMassRel}
\eea
Then, in the series expansion of $E_{\rm tot}(r)$ in $\alfs$,
the leading IR renormalon contained in $c_0$ is canceled
against the leading IR renormalon contained in $m_{\rm pole}$,
which is estimated from the diagrams in 
Fig.~\ref{Pot-SE-BubbleChain}(b)
\cite{Pineda:id,Hoang:1998nz,Beneke:1998rk}.
\begin{figure}
\begin{center}
\vspace*{-0.5cm}
\includegraphics[width=12cm]{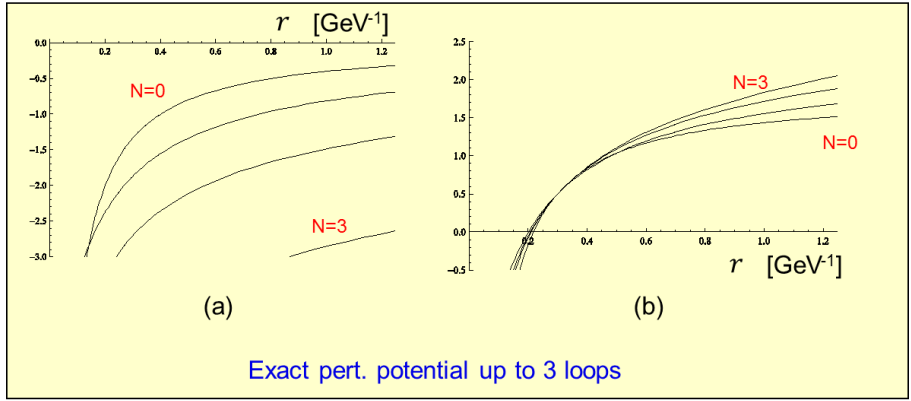}
\end{center}
\vspace*{-.5cm}
\caption{\small
Comparisons of perturbative predictions of $V_{\rm QCD}(r)$
up to ${\cal O}(\alfs^{N+1})$
for $N=0$,\,1,\,2 and 3, as they are [(a)],
and after the leading renormalon is canceled by
adding an $r$-independent constant
to each potential such that it takes a common value at 
$r=0.3~{\rm GeV}^{-1}$ [(b)].
\label{PotUpto3Loops}}
\end{figure}
The remaining largest renormalon of $E_{\rm tot}(r)$ becomes 
order $\LQ^3r^2$ in $c_2\, r^2$.
As a result, uncertainties of the perturbative series for
$E_{\rm tot}(r)$ are much more suppressed at $r\ll\LQ^{-1}$
as compared to those for $V_{\rm QCD}(r)$, namely,
the perturbative series for the former is much more
convergent than that for the latter.
This feature is shown in Figs.~\ref{PotUpto3Loops}(a)(b).
\pagebreak
In the left figure the perturbative predictions for
$V_{\rm QCD}(r)$ up to ${\cal O}(\alfs^{N+1})$ are shown
($N=0$,\,1,\,2 and 3).
As we increase the order, there are large, almost
constant shifts downwards, which make convergence of the perturbative series 
very poor in the displayed range of $r$.
In the right figure, we effectively cancel the
renormalon in $c_0$ by adding an $r$-independent constant
to each potential such that it takes a common value at 
$r=0.3~{\rm GeV}^{-1}$.\footnote{
Since $d_4$ of eq.~(\ref{PoleMSbarMassRel}) 
is unknown at present, we cannot display 
$E_{\rm tot}(r)$ at ${\cal O}(\alfs^{4})$.
We observe a qualitatively similar feature
as in Fig.~\ref{PotUpto3Loops}(b)
if we plot $E_{\rm tot}(r)$ up to ${\cal O}(\alfs^{3})$.
}
As can be seen, the predictions in Fig.~\ref{PotUpto3Loops}(b)
is much more convergent.
Furthermore, we note that the potential becomes steeper
as we include higher-order corrections \cite{Sumino:2001eh}.
This is a desirable tendency for the potential to approach
the correct shape by inclusion of higher-order corrections.
If we estimate the perturbative series of the
potential by the bubble-chain
contributions and make similar plots, they look qualitatively 
the same as
Figs.~\ref{PotUpto3Loops}(a)(b). 
Thus, the renormalon estimates
are indeed good approximations for the static potential.

\begin{figure}
\begin{center}
\vspace*{.5cm}
\includegraphics[width=11cm]{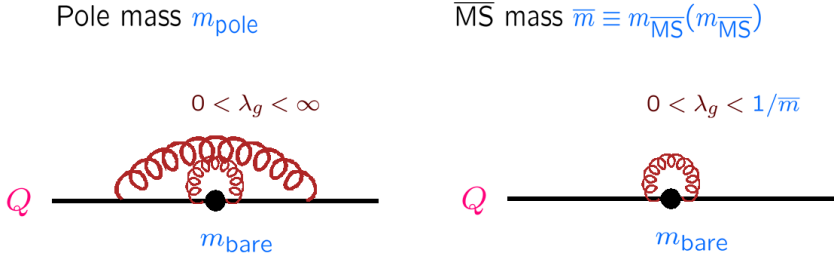}
\vspace*{-.5cm}
\end{center}
\caption{\small
Schematic diagrams representing intuitive pictures of the
pole mass and {\msbar} mass of a quark.
\label{Pole-MSbar-Masses}}
\end{figure}
Let us explain physical interpretations of the quark pole mass
and {\msbar} mass.
See Fig.~\ref{Pole-MSbar-Masses}.
The pole mass $m_{\rm pole}$ is defined as the position of the pole of a
quark full propagator defined in perturbation theory.
It is equivalent to the self-energy of a quark in its rest frame.
Since a quark has color, gluons with arbitrarily long
wave-lengths (small momenta) can couple to the quark
and contribute to the self-energy.
On the other hand, the {\msbar} mass 
$\overline{m}$
is defined as the contributions to the quark self-energy from
gluons with wave-lengths smaller than $1/\overline{m}$.
Hence, the difference between $m_{\rm pole}$ and $\overline{m}$
is given by the contributions to the 
quark self-energy from IR gluons
(wave-lengths larger than $1/\overline{m}$).
Twice of this 
difference corresponds to the additional terms in $E_{\rm tot}(r)$
[twice of eq.~(\ref{PoleMSbarMassRel})
minus $2\overline{m}$].

\begin{figure}[t]\centering
\includegraphics[width=13cm]{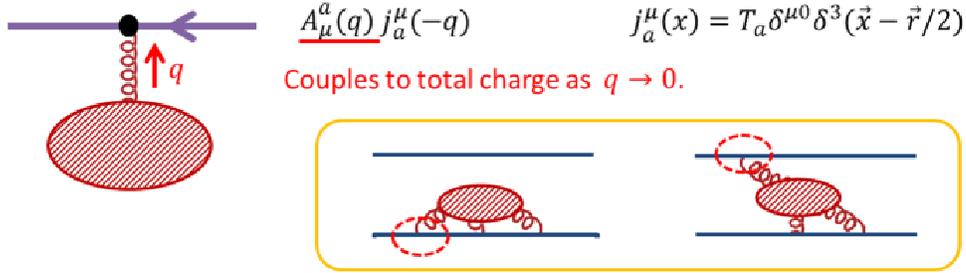}
\caption{\small
As a general feature of gauge theory, a gluon, which couples to static 
currents $j^{\mu}_a \propto \delta^{\mu 0}$,
couples to the total charge of the system in the IR limit, $q\to 0$.
Diagrammatically both self-energy and potential-energy type diagrams
are needed for realizing this feature, hence,
for a color-singlet system, a cancellation takes place
between the two types of diagrams.
\label{FigIRgluon}
}
\end{figure}

The cancellation of IR contributions in $E_{\rm tot}(r)$
is a general property of gauge theory, which holds
beyond the estimates by the bubble-chain diagrams.
In fact, a gluon, which couples to static 
currents $j^{\mu}_a \propto \delta^{\mu 0}$
via minimal coupling $A_\mu^a\, j^\mu_a$, couples to
the total charge of the system 
$Q_a^{\rm tot}=\sum_{i} j^0_{a,i}(q=0)$ ($i=Q,\bar{Q}$)
in the zero momentum limit $q\to 0$.
Namely, an 
IR gluon decouples
from the color-singlet system.
Diagrammatically an IR gluon observes the total
charge of the system when both self-energy diagrams\footnote{
In the large mass limit contributions from IR region
to the pole mass approximate 
IR contributions to the self-energy 
of a static charge.
} 
and potential-energy diagrams
are taken into account.
This means that 
a cancellation takes place between these two types of diagrams,
see Fig.~\ref{FigIRgluon}.
In perturbative QCD, convergence of perturbative series
become worse as contributions from IR gluons grow.
Oppositely, after cancellation of IR contributions,
convergence of perturbative predictions improve.
This can be considered as a general property of a gauge theory
which is strongly interacting at IR.

Thus, the dominant IR contributions to $V_{\rm QCD}(r)$ are contained in 
the $c_0$ and $c_2\,r^2$ terms 
in Eq.~(\ref{naive-expansion}).
The renormalon in $c_0$ is canceled against 
the pole mass in $E_{\rm tot}(r)$.
On the other hand, the renormalon in $c_2\,r^2$
is replaced by a non-perturbative matrix element
in OPE of $V_{\rm QCD}(r)$ 
within an EFT ``potential-NRQCD''
\cite{Brambilla:1999qa}, as explained below.

The potential-NRQCD \cite{Brambilla:2004jw} is an EFT for describing 
interactions between IR gluons 
and heavy quark-antiquark ($Q\bar{Q}$) bound-states.
Since $Q$ and $\bar{Q}$ are heavy, the size $r$ of a bound-state is small
$r< \LQ^{-1}$.
\begin{figure}[t]\centering
\includegraphics[width=6cm]{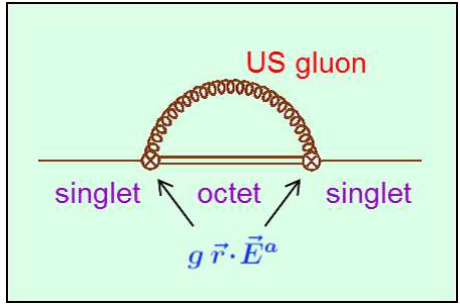}
\caption{\small
Diagram in potential-NRQCD EFT which contributes to 
$E_{\rm tot}(r)$.
The single and double straight lines represent the singlet and
octet $Q\bar{Q}$ states, respectively.
``Ultra-soft (US) gluon'' represents an IR gluon
with energy appropriate for transitions between
different $Q\bar{Q}$ states.
\label{UScorr-pNRQCD}
}
\end{figure}
In this EFT, the leading interaction between a color-singlet
$Q\bar{Q}$ state and IR gluon in multipole expansion 
in $\vec{r}$ is 
a dipole-type interaction $\vec{r}\cdot \vec{E}^a$,
where the $r^0$ coupling with the total charge
vanishes due to the color-singlet nature of the
$Q\bar{Q}$ state.
Hence, the leading contribution to the energy
of $Q\bar{Q}$ is expressed in terms of a matrix element\footnote{
More precisely the non-perturbative contribution is given as
a non-local gluon condensate
\bea
V_{\rm US}(r)=
- \frac{i g_S^2}{6}
\int_0^\infty \! \! dt \, e^{-i \, \Delta V(r)\, t} \,
\bra{0} \vec{r}\!\cdot\!\vec{E}^a(t) \, \varphi_{\rm adj}(t,0)^{ab}\,
\vec{r}\!\cdot\!\vec{E}^b(0) \ket{0},
\eea
where $\Delta V(r)$ denotes the
difference between the octet and singlet potentials,
and $\varphi_{\rm adj}(t,0)^{ab}$ represents a color flux
spanned between the coordinates $(0,\vec{0})$ and $(t,\vec{0})$.
$V_{\rm US}(r)$ reduces to the local gluon condensate
$\sim r^3\langle G_{\mu\nu}^a(0)^2 \rangle$
at very small $r$
($\ll \Delta V(r)^{-1}$), which lies outside the range of
$r$ considered here.
See \cite{Brambilla:1999qa} for details.
}
\bea
\langle \vec{r}\cdot \vec{E}^a(t) \, \vec{r}\cdot \vec{E}^b(0) \rangle .
\eea
It contains the dipole interaction twice, since
the gluon emitted from the $Q\bar{Q}$ bound-state needs to be
reabsorbed; see Fig.~\ref{UScorr-pNRQCD}.
Thus, it is proportional to $r^2$ and in fact has a
form which exactly matches to replace the order $\LQ^3r^2$ renormalon
in $c_2\,r^2$ \cite{Brambilla:1999qa,Sumino:2004ht}.
This exemplifies the relation between
IR renormalons and non-perturbative matrix elements as explained in 
Sec.~3.2.

Note that there are no IR contributions to the 
${c_{-1}}{r}^{-1}$ and $c_1r$ terms in Eq.~(\ref{naive-expansion}) 
by IR renormalons or by non-perturbative
matrix elements in OPE of potential-NRQCD.
Then, it  is logically expected that the
${c_{-1}}{r}^{-1}$ and $c_1r$ terms are
dominated by UV contributions.\footnote{
After canceling the renomalons, the
$r$ independent term of $E_{\rm tot}(r)$ is
also dominated by UV contribution.
}
In the next subsection, we answer to the question
``What are the UV contributions to $V_{\rm QCD}(r)$\,?''

\subsection{UV contributions as a ``Coulomb+linear'' potential}

In this subsection we show that the UV contribution to $V_{\rm QCD}(r)$
in perturbative QCD takes a 
``Coulomb+linear'' form,
after resummation of logarithms 
\cite{Sumino:2003yp,Sumino:2004ht,Sumino:2005cq}.
Namely, the perturbative prediction can be
cast into the form
\bea
&&
V_{\rm QCD}(r)={V_C(r)} + {\rm const}.+
{\sigma\,r}
+{{\cal O}(\LQ^3r^2)}
~~~{\rm at}~~r\simlt \LQ^{-1} .
\label{VasC+L}
\\&&
~~~~~~~~~~~~~~~~~\mathlarger{\scriptstyle r^{-1}}
~~~~~~~\mathlarger{\scriptstyle r^0}~~~~~~~~\,\mathlarger{\scriptstyle r^1}
~~~~~~~~~~\mathlarger{\scriptstyle r^2}
\nonumber
\eea
We first show the results and afterwards explain how they can be derived.

\begin{figure}
\begin{center}
\begin{tabular}{cc}
\includegraphics[width=7cm]{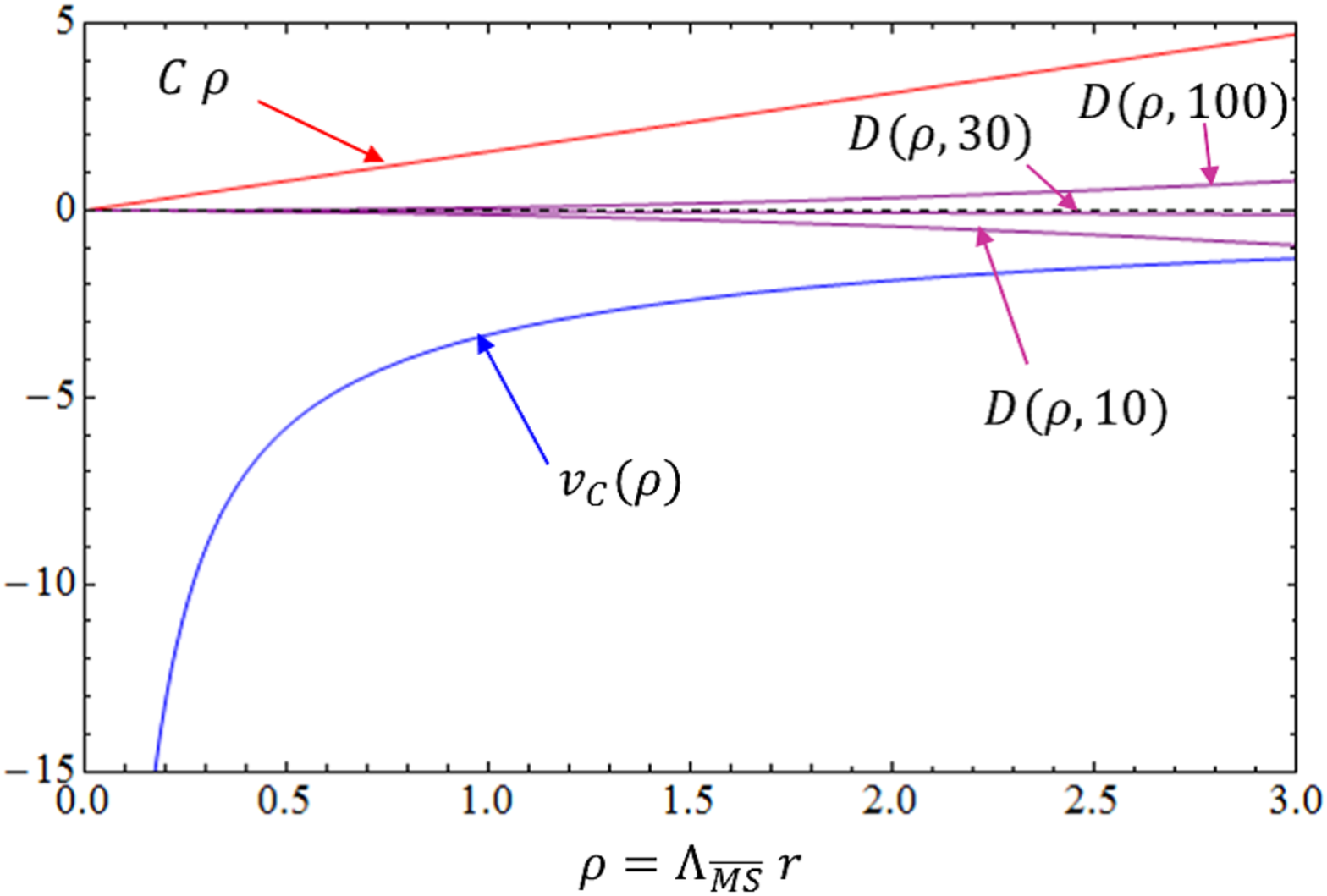}
&
\includegraphics[width=8cm]{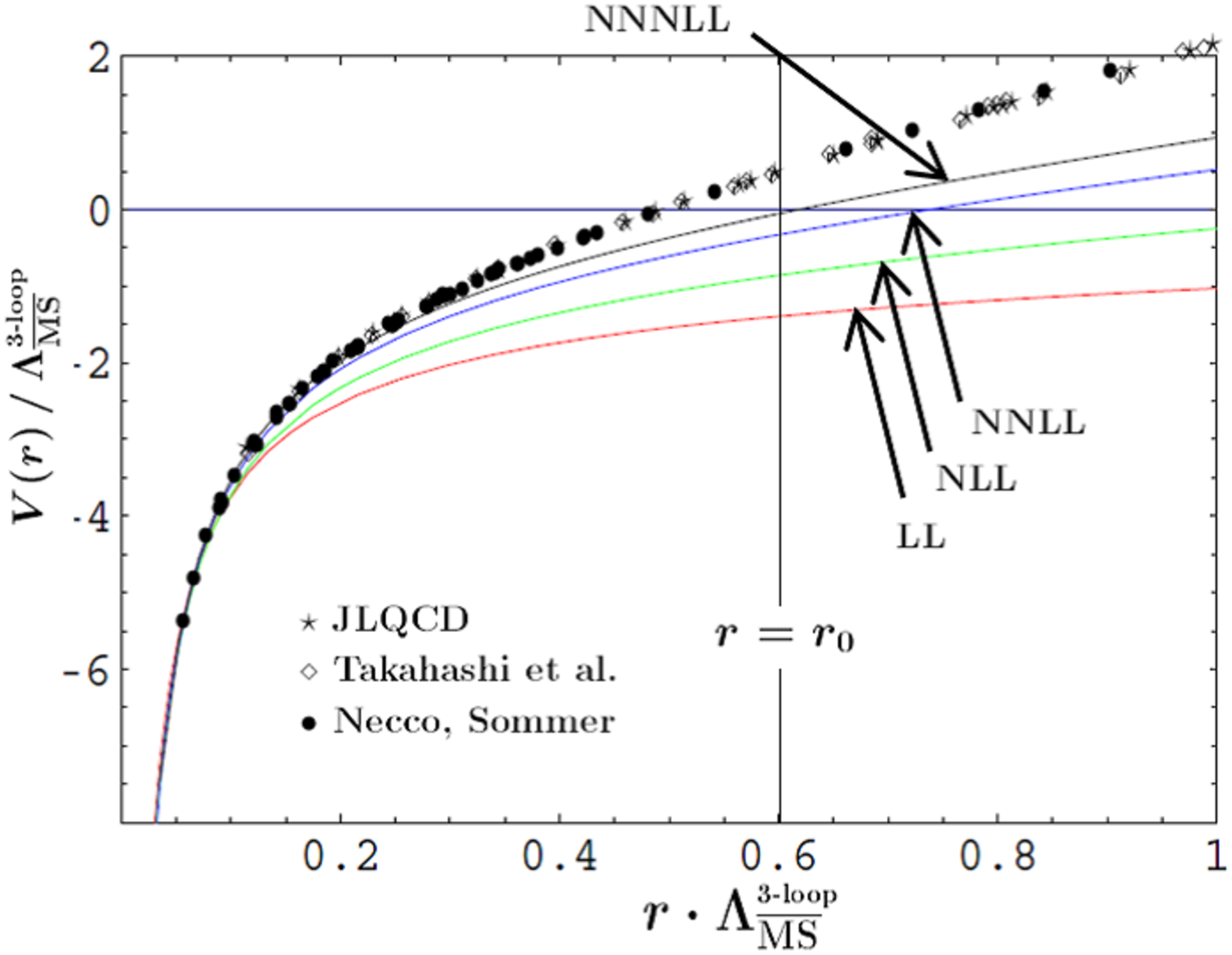}\\
(a)&~~~~~~(b)
\end{tabular}
\end{center}
\vspace*{-.5cm}
\caption{\small
(a)
``Coulomb,'' linear and ${\cal O}(\LQ^3r^2)$ parts of
Eq.~(\ref{VasC+L}) in the LL approximation.
See \cite{Sumino:2003yp,Sumino:2005cq} for details.
(b) 
Comparison of lattice computations
\cite{Necco:2001xg,Takahashi:2002bw,Aoki:2002uc}
of $V_{\rm QCD}(r)$ and ${V_C(r)} + {\sigma\,r}$.
Solid lines represent ${V_C(r)} + {\sigma\,r}$,
defined by Eqs.~(\ref{formula-Vc}) and (\ref{formula-sigma}),
at different orders of log resummations.
\label{LogResumPot}}
\end{figure}
Fig.~\ref{LogResumPot}(a) shows each part of 
the above decomposition of the potential in the case that
leading-logarithms (LL) are resummed.
$v_C(\rho)$ and $C\rho$ correspond to the ``Coulomb'' and linear
terms, respectively.
Differences among
$D(\rho,N)$'s represent the level of uncertainties induced by
the ${{\cal O}(\LQ^3r^2)}$ renormalon.
Here and hereafter, the $r$-independent constant is neglected.
The figure shows that the LL potential is approximated well by the
``Coulomb+linear'' potential ${V_C(r)} + {\sigma\,r}$
in the range $r\simlt \Lambda_{\overline{\rm MS}}^{-1}$, 
since the ${{\cal O}(\LQ^3r^2)}$
uncertainties are not significant.

Fig.~\ref{LogResumPot}(b) shows a
comparison of lattice data
and ${V_C(r)} + {\sigma\,r}$ in different orders
of log resummations.
As can be seen, with increasing  order, the range where
the perturbative prediction agrees with the lattice data extends
to larger $r$.
These predictions for ${V_C(r)} + {\sigma\,r}$ are expressed
by the parameters of perturbative QCD.
For instance, the coefficient of the linear potential 
at NLL is given by
\bea
\sigma_{\rm NLL} = \frac{2\pi C_F}{\beta_0} \,
\Bigl( \Lambda_{\overline{\rm MS}}^{\mbox{\scriptsize 2-loop}} \Bigr)^2
\, \frac{e^{-\delta}}{\Gamma (1+\delta)} \,
\biggl[ 1 + \frac{a_1}{\beta_0} \, \delta^{-1-\delta} \, e^\delta \,
\gamma (1+\delta,\delta) \biggr] ,
\eea
where $C_F=4/3$ is the Casimir
operator for the fundamental representation, and
$\delta=\beta_1/\beta_0^2$; see  
\cite{Sumino:2003yp,Sumino:2005cq}
for details.

\begin{figure}
\begin{center}
\begin{tabular}{cc}
\includegraphics[width=5cm]{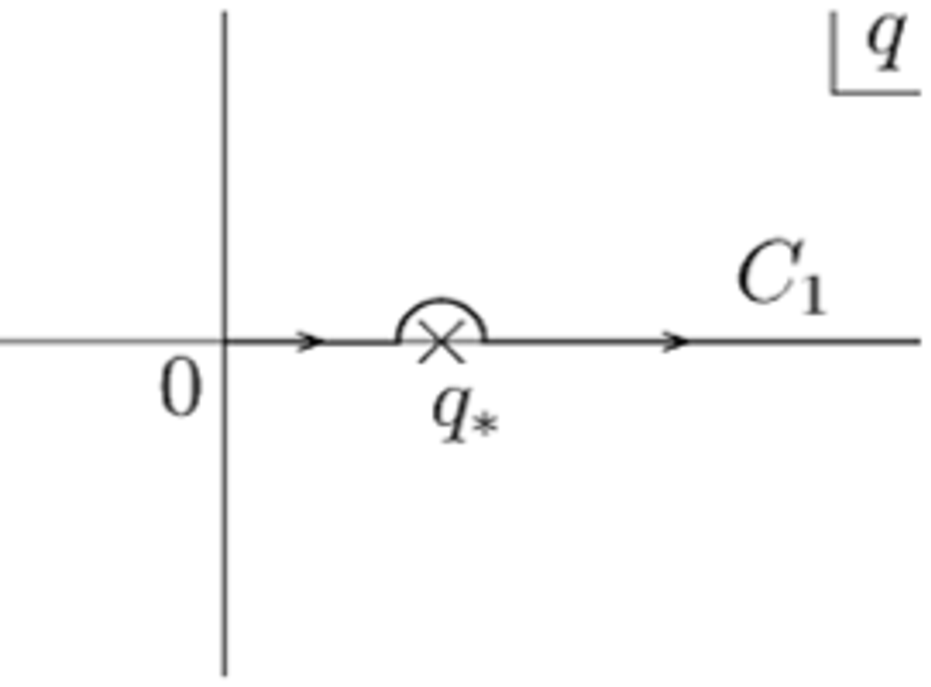}
&
\includegraphics[width=6.2cm]{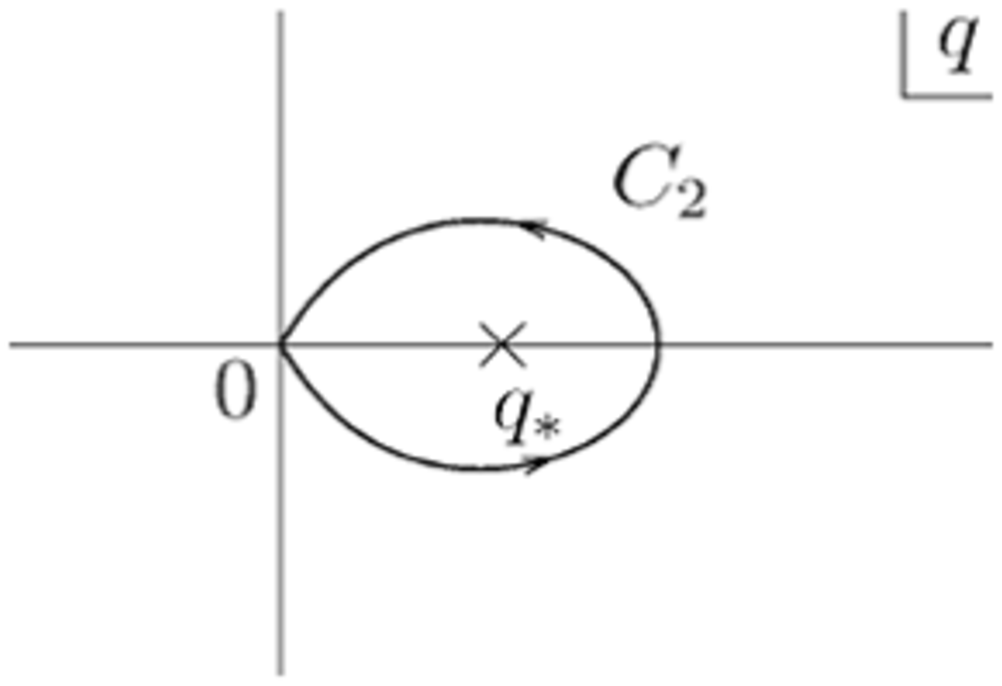}
\end{tabular}
\end{center}
\vspace*{-.5cm}
\caption{\small
The contours $C_1$ and $C_2$ of integrals in Eqs.~(\ref{formula-Vc})
and (\ref{formula-sigma}).
\label{C1C2}}
\end{figure}

The formulas for ${V_C(r)}$ and ${\sigma\,r}$ are given as contour
integrals in the complex $q$-plane as follows:
\bea
&&
V_C(r) = - \frac{C_F}{\pi i}
\int_{C_2}\! dq \, \frac{\alpha_V(q) }{qr}
- \frac{2C_F}{\pi} \, {\rm Im}
\int_{C_1}\! dq \, \frac{e^{iqr}}{qr} \, \alpha_V(q) ,
\label{formula-Vc}
\\&&
{\sigma} = \frac{C_F}{2\pi i} \int_{C_2}\! dq \, q \, \alpha_V(q) 
.
\label{formula-sigma}
\eea
The contours $C_1$ and $C_2$ are shown in Fig.~\ref{C1C2}.
Here, the 
$V$-scheme coupling constant $\alpha_V(q)$ is defined from
the Fourier transform of the potential by
\bea
V_{\rm QCD}(r) = \int{\frac{d^3\vec{q}}{(2\pi)^3}}
\left[-4\pi C_F\frac{\alpha_V(q)}{q^2}\right] e^{i\vec{q}\cdot\vec{r}}
~~~;~~~
q=|\vec{q}|,
\eea
and can be computed in perturbative QCD.

In the case of LL approximation,
\bea
\alpha_V(q)=\frac{2\pi}{\beta_0\log(q/\Lambda_{\overline{\rm MS}}^{\mbox{\scriptsize 1-loop}})}
\eea
coincides with the one-loop running coupling constant.
The ``Coulomb'' potential ${V_C(r)}$ is easily computed numerically
using the above formula, while its asymptotic behavior can be
extracted analytically:
\bea
V_C^{\rm (LL)}(r)
\to
\left\{
\begin{array}{l}\displaystyle
-\frac{2\pi C_F}{\beta_0 }\,
\frac{1}{r \Bigl|\log (
\Lambda_{\overline{\rm MS}}^{\mbox{\scriptsize 1-loop}}
\,r)\Bigr|}
,
~~~~~~~~~~\,
r \to 0 ,
\\ \displaystyle
\rule{0mm}{10mm}
 - \frac{4\pi C_F}{\beta_0 r} ,
~~~~~~~~~~
~~~~~~~~~~
~~~~~~~~~
r \to \infty .
\end{array}
\right.
\eea
At short-distance it tends to a Coulomb potential with
the correct logarithmic correction as determined by the RG equation;
at large-distance, it approaches a Coulomb potential;
see Fig.~\ref{LogResumPot}(a).
On the other hand, the coefficient of the linear potential
can be computed analytically using the Cauchy theorem:
\bea
\sigma_{\rm LL} = \frac{2\pi C_F}{\beta_0} \,
\Bigl( \Lambda_{\overline{\rm MS}}^{\mbox{\scriptsize 1-loop}} \Bigr)^2 .
\eea

To clarify the nature of ${V_C(r)} + {\sigma\,r}$,
we compare it with the UV contribution to $V_{\rm QCD}(r)$,
defined as a Wilson coefficient in potential-NRQCD.
Note that this EFT is valid in the case $r^{-1}\gg\mu_f\gg\LQ$,
where physical modes above the factorization scale $\mu_f$ have been
integrated out.
The relevant Wilson coefficient can be defined 
(in a hard cut-off scheme)
as
\bea
V_{\rm UV}(r;\mu_f) = \int_{q>\mu_f}{\frac{d^3\vec{q}}{(2\pi)^3}}
\left[-4\pi C_F\frac{\alpha_V(q)}{q^2}\right] e^{i\vec{q}\cdot\vec{r}}
.
\label{VUV}
\eea
Since $\mu_f\gg\LQ$,
within the integral region $q>\mu_f$, 
$\alpha_V(q)$ can be computed accurately
in perturbative QCD.
In particular, $V_{\rm UV}$ is shown to be
free of IR renormalons (in the estimate by the bubble-chain diagrams).
$V_{\rm UV}$ represents 
the leading UV contribution to the static potential.\footnote{
In  OPE in the form of Eq.~(\ref{OPE}), $P$ should be identified
with $r^{-1}$, and the matrix element multiplying the
Wilson coefficient $V_{\rm UV}$
is $1=\braket{0}{0}=\bra{S}S^\dagger S\ket{S}$,
where $S$ and $\ket{S}$ denote the field operator
and the state vector
of the color-singlet bound-state, respectively.
}

\begin{center}
\begin{tabular}{|p{15.3cm}|}
\hline
~~
It can be proven that
\bea
V_{\rm UV}(r;\mu_f) - \Bigl[ V_C(r) + \sigma r \Bigr] = 
{\rm const.}+{\cal O}(\mu_f^3 r^2) .
\label{ImportantRel}
\eea
This shows that, in perturbative QCD, the ``Coulomb'' and linear parts of
$V_{\rm QCD}(r)$ are determined 
by UV contributions and independent of the factorization scale
$\mu_f$.
\\
\hline
\end{tabular}
\end{center}

${V_C(r)} + {\sigma\,r}$ represents the 
``Coulomb'' and linear parts of
$V_{\rm QCD}(r)$ in perturbative QCD by Eq.~(\ref{VasC+L}),
and $V_{\rm UV}$ represents the UV contribution
to $V_{\rm QCD}(r)$.
The above relation shows that there are no Coulomb or
linear potential in the difference
(at $r\ll \mu_f^{-1}$), namely
they are included in the UV contribution.
It is probably not surprising that the ``Coulomb'' potential
arises from the UV contribution, since singular behavior
as $r\to 0$ can only stem from the short-wave length
modes $\lambda < r$
of gluons exchanged between the
color-singlet pair $Q$ and $\bar{Q}$.
It would be surprising, however, that even the
linear potential is a short-distance contribution.
The above relation also shows that, since
${V_C(r)} + {\sigma\,r}$ is independent of $\mu_f$,
the `Coulomb'' and linear potentials contained in $V_{\rm UV}$ are
insensitive to the IR cut-off of eq.~(\ref{VUV}).
In this sense, they are genuinely UV in nature.
It is consistent, since if we raise $\mu_f$ in $V_{\rm UV}$,
$\mu_f$-dependent part  (an IR part) 
of $V_{\rm UV}$ should be compensated by
IR operators in the EFT, but
there are no IR operators which
can absorb the $r$-dependences
of ${V_C(r)}$ and ${\sigma\,r}$.

The proof of Eq.~(\ref{ImportantRel}) is given in the
Appendix, which is fairly easy to understand.
An essential point of the proof is that contributions from the IR
scale of order $\LQ$ are given as contour integrals
near the singularity of $\alpha_V(q)$ which arises as a result of
resummation of logarithms.
We can then separate the IR contributions from $V_{\rm QCD}(r)$.
This is in parallel to the argument given in Sec.~3.4,
that 
contributions from different scales can be separated
as contour
integrals surrounding the corresponding singularities
in a Feynman diagram.
The difference is that at each order of perturbation 
we cannot see a singularity
corresponding to the scale $\LQ$, as it arises only after
resummation of logarithms.
Nevertheless, it should be noted that, as we saw in Sec.~3.1,
inclusion of the higher-order corrections approximate to the
resummed contribution, and sensitivity to the $\LQ$ scale
becomes apparent gradually in perturbation theory
even without a log resummation.

At this stage we stress the need for
the IR renormalization of the 
Wilson coefficient $V_{\rm UV}(r)$.
Eq.~(\ref{VUV}) corresponds to introducing a hard cut-off
to define a Wilsonian EFT.
On the other hand, as we explained, the modern way to define
an EFT is to separate scales by using asymptotic expansion
in dimensional regularization.
In this case, separation of IR contributions is non-trivial,
since the IR scale $\LQ$ does not appear at any fixed-order
of the perturbative expansion.
In fact, one can show that the bare Wilson coefficient $V_{\rm UV}(r)$
defined in dimensional regularization in EFT
coincides with
$V_{\rm QCD}(r)$ in perturbative QCD to all orders
in perturbative expansion.
Therefore, this $V_{\rm UV}(r)$ includes uncertainties
by IR renormalons.
It means that one needs to explicitly renormalize $V_{\rm UV}(r)$
at IR by subtracting IR contributions in some prescription.\footnote{
This is a demerit of using
asymptotic expansion for scale separation in perturbative
QCD.
It is not needed in construction of an EFT
using a hard cut-off.
}
One way is to subtract an estimate of contributions from
IR renormalons at each order of perturbative expansion
\cite{Pineda:2002se}.
The other method is the one we advocate above \cite{Sumino:2005cq}:
resum logarithms first, identify the IR singularity 
corresponding to the $\LQ$ scale in the
resummed contribution, and separate its contribution by a
prescription similar to the asymptotic expansion;
in this way we can subtract the IR sensitive parts and
renormalize them into the pole mass and non-perturbative
matrix element;
as a result we obtain ${V_C(r)}+{\sigma\,r}$ as a
renormalized Wilson coefficient.
Both prescriptions
lead to similar results numerically, but the consequences
are probably easier to interpret in the latter method.

There are two ways to derive the formulas
for ${V_C(r)}$ and ${\sigma\,r}$,
Eqs.~(\ref{formula-Vc}) and (\ref{formula-sigma}).
One way is along the line of proof in the Appendix.
In this way, ${V_C(r)}+{\sigma\,r}$ can be
related to the perturbative evaluation of
the Wilson coefficient $V_{\rm UV}$ after log resummation.
The other way is to extract ${V_C(r)}+{\sigma\,r}$
from the perturbative series of $V_{\rm QCD}(r)$.
It can be shown that, while the perturbative series is
converging (i.e., for $n<n_0$ of Sec.~3.1), the sum of the
series approaches ${V_C(r)}+{\sigma\,r}$
up to an order $\LQ^3r^2$ uncertainty.

\subsection{Implication and physical interpretation}
\clfn

Using ${V_C(r)}+{\sigma\,r}$ thus obtained,
we can compute the total energy of a static
$Q\bar{Q}$ pair as
\bea
E_{\rm tot}(r)=2\overline{m}+{\rm const.}+
{V_C(r)}+{\sigma\,r}+{\cal O}(\LQ^3r^2) ,
\eea
where the $r$-independent part is also
UV dominant and accurately predictable.
In principle,
the ${\cal O}(\LQ^3r^2)$ uncertainty can be replaced by
the non-perturbative matrix element in OPE.
However, there has
been no direct evaluation of the matrix element
so far; it is only estimated to be small from a comparison 
of the potential to
lattice data; see Figs.~\ref{3LoopQCDPot} and \ref{LogResumPot}(b).
The spectrum of a heavy quarkonium system,
such as bottomonium or (would-be) toponium,
can be computed roughly as the energy eigenvalues
of the quantum mechanical Hamiltonian
\bea
H=\frac{\vec{p}^{\,2}}{2m_{\rm pole}}+E_{\rm tot}(r) .
\label{LOHamiltonian}
\eea
(More accurate prediction is possible using the
potential-NRQCD framework.
Currently
the spectrum is known up to NNNLO \cite{Kiyo:2013aea}.)

\begin{figure}[t]\centering
\vspace*{5mm}
\includegraphics[width=6cm]{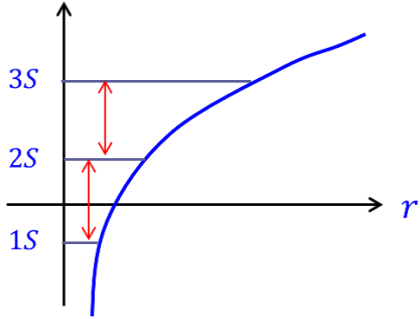}
\caption{\small
A schematic diagram showing the energy levels of
the Hamiltonian Eq.~(\ref{LOHamiltonian}).
The level spacings are order $\LQ(\LQ/m)^{1/3}$ if
they are predominantly determined by the linear part
$\sigma r$ of the potential.
\label{Levels-in-C+L_Pot}
}
\end{figure}
A linear potential of order $\LQ^2\,r$ generates
level spacings between different $S$-states
of order $\LQ(\LQ/m)^{1/3}$.
On the other hand, a Coulomb potential $\sim -\alpha_s/r$ generates
level spacings of
order $\alpha_s^2 m$.
For the bottomonium states, the linear potential
is estimated to be comparable to or
more important than the Coulomb potential
in generating these level spacings,
whereas for the toponium states, the 
Coulomb potential by far dominates over
the linear potential.
Thus, a major part of the perturbative QCD predictions for
the level spacings between different bottomonium $S$-states
is order $\LQ(\LQ/m)^{1/3}$.
See Fig.~\ref{Levels-in-C+L_Pot}.

We may develop a microscopic understanding on 
the composition of the energy
of a bottomonium state based on perturbative QCD.
According to the discussion in Sec.~4.1,  
infrared gluons decouple
in the computation of the energy of a  bottomonium state $X$.
\begin{figure}[t]
\vspace*{0.5cm}
\begin{center}
    \includegraphics[width=12cm]{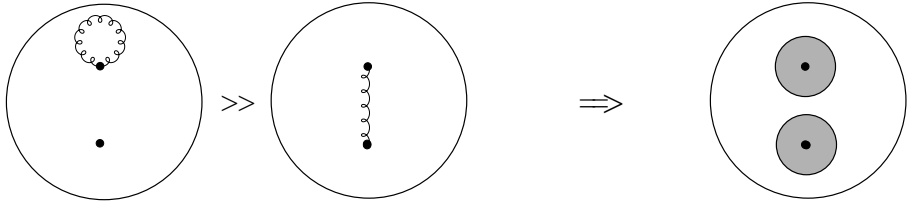}
\end{center}
\caption{\small
The total energy of a heavy quarkonium state is carried by the
$\overline{\rm MS}$ masses of quark and antiquark and by 
the gluons inside the bound-state.
In the latter contributions the self-energies of quark and antiquark 
dominate over the potential energy between the two particles.
\label{physpic}}
\end{figure}
The energy consists of the self-energies
of $b$ and $\bar{b}$ and the potential energy between $b$ and $\bar{b}$,
where gluons whose wave-lengths are smaller than the bound-state
size $a_X$ contribute.
At IR the sum of the self-energies and the potential energy cancel.
\pagebreak
On the other hand, at UV, the potential energy quickly dumps
due to the rapid oscillation factor $e^{iqr}$ for large $q$
in the potential energy.
It means that
the major contribution 
to the bottmonium energy comes from the region (in momentum space)
$1/a_X$ $ \simlt$ $ q$ $\simlt$ $\overline{m}_b$ of the self-energy corrections 
of $b$ and $\bar{b}$,
apart from the constant contribution $2 \overline{m}_b$.
See Fig.~\ref{physpic}.

\begin{figure}[h]
\begin{center}
\begin{tabular}{cc}
\includegraphics[width=8cm]{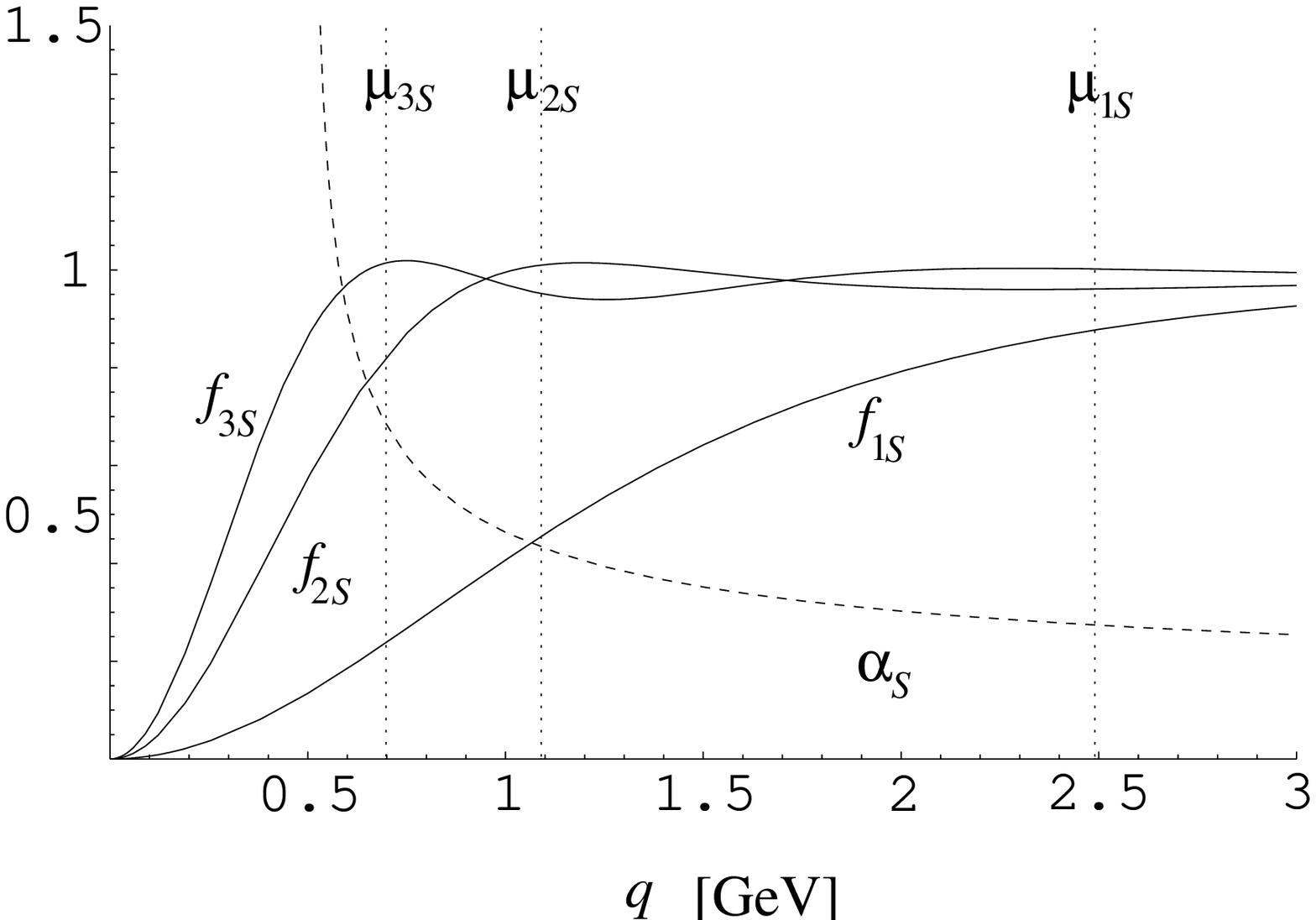}~~
&
\includegraphics[width=5.5cm]{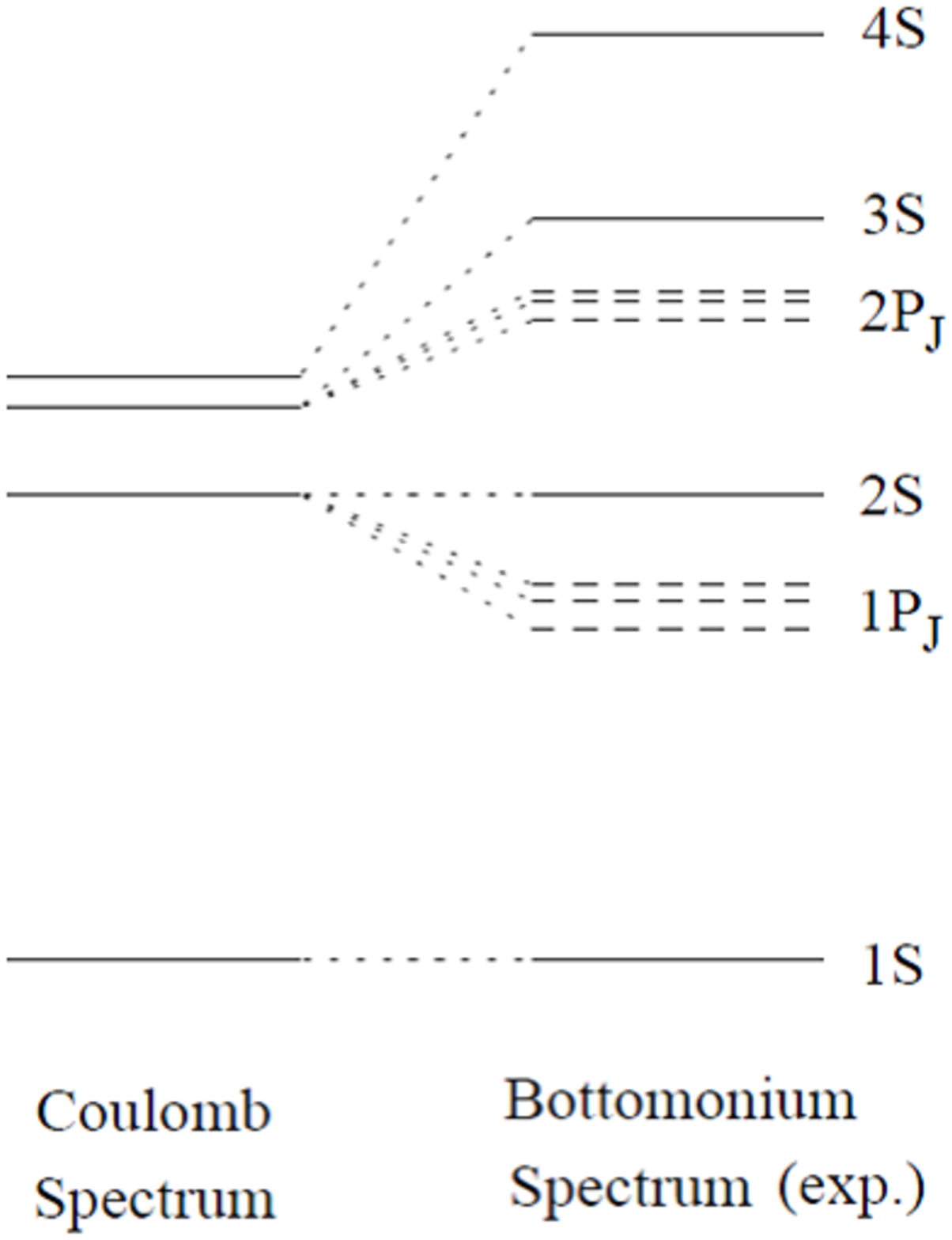}\\
(a)~~&~(b)
\end{tabular}
\end{center}
\caption{\small
(a) The support functions $f_X(q)$ used to express the
energy of the bottomonium state $X$ in Eq.~(\ref{eth3})
for $X=1S$, $2S$ and $3S$ 
\cite{Brambilla:2001fw}.
The running coupling constant $\alpha_s(q)$ 
close to the dumping scale of $f_X(q)$ grows rapidly
as $X$ varies from the $1S$ to $3S$ states.
(b)~Comparison of the Coulomb spectrum and observed
bottomonium spectrum.
The Coulomb spectrum is scaled such that the $1S$--$2S$ level
spacing coincides with that of the bottomonium spectrum.
\label{BottomoniumEnergy}}
\vspace*{.5cm}
\end{figure}

In fact, the composition of the energy in momentum space 
can be expressed approximately as \cite{Brambilla:2001fw}
\bea
E_X &\simeq& 2 \overline{m}_b + \frac{2C_F}{\pi} \int_0^{\overline{m}_b} 
\!\! dq \, \alpha_s (q) 
\, f_X(q),  
\label{eth3}
\eea
where $f_X(q)$ is a support function constructed
from the wave-function of the bound-state $X$, which is roughly
unity in the region 
$1/a_X$ $ \simlt$ $ q$ $\simlt$ $\overline{m}_b$;
see Fig.~\ref{BottomoniumEnergy}(a).

A characteristic feature of the bottomonium spectrum
in comparison to the Coulomb spectrum is that
the level spacings among the bottomonium
excited states are much wider than those of the Coulomb spectrum.
The level spacings of the Coulomb spectrum 
decrease quickly for higher levels.
The difference from the
Coulomb spectrum results from the linear rise of the potential.
See Figs.~\ref{Levels-in-C+L_Pot} and \ref{BottomoniumEnergy}(b).

The size $a_X$ of the state $X$ becomes larger
for higher excited states.
Then gluons with longer wave-lengths can contribute
to the energy of $X$.
Positive contributions to the self-energies increase
rapidly since interactions of IR gluons become
stronger by the running of the coupling constant.
In Fig.~\ref{BottomoniumEnergy}(a) also $\alpha_s(q)$ is shown. 
We see that as the state varies from
$X= 1S$ to $3S$, the coupling $\alpha_s(q)$, close to the 
dumping scale of $f_X(q)$, grows rapidly.
According to Eq.~(\ref{eth3}), as the integral region 
extends down to smaller $q$, the self-energy contributions grow
rapidly in comparison to the non-running case.
(Note that the non-running case corresponds to the
Coulomb spectrum.)
The self-energies push up the energy levels of the excited states considerably
and widen the level spacings among the excited states
as compared to the Coulomb case.

\pagebreak
Hence, we may draw the following 
qualitative pictures for the energies of the bottomonium states
\cite{Brambilla:2001fw}:
\begin{itemize}
\item[(I)]
The energy of a bottomonium state mainly  consists of
(i) the $\overline{\rm MS}$ masses of $b$ and $\bar{b}$, and 
(ii) contributions to the self-energies of $b$ and $\bar{b}$
from gluons with wavelengths $1/\overline{m} \simlt \lambda \simlt a_X$.
The latter contributions may be regarded as the difference between
the (state-dependent) constituent quark masses and the current quark masses.
\item[(II)]
The energy levels between excited states are widely separated
as compared to the Coulomb spectrum. 
This is because the self-energy contributions (from
$1/\overline{m} \simlt \lambda$ $\simlt a_X$)
grow rapidly as the physical size $a_X$ of the bound-state increases.
\end{itemize}
We conjecture that 
the conventional picture, that
the mass of a light hadron consists of the constituent quark masses,
can be viewed as an extrapolation of picture (I),
although it lies outside the validity range of perturbative QCD.

The predictions of the bottomonium spectrum in
perturbative QCD  and their detailed quantitative analyses 
can be found in \cite{Kiyo:2013aea}.

\pagebreak
\subsection{Lessons drawn from the analysis of heavy quarkonium states}

There are lessons that can be drawn from the
analysis of the heavy quarkonium states and static potential, which may be
applicable in more general contexts.
\begin{itemize}
\item[(1)] 
One should carefully examine, from which power of 
$\LQ=\mu\exp\Bigl[-\frac{2\pi}{\beta_0\alpha_s(\mu)}\Bigr]$ 
non-perturbative contributions start,
and to which extent perturbative QCD is predictable
(as we approach from the short-distance region).
\end{itemize}
It is a priori not obvious that the $\LQ^2\,r$ potential is
dominated by UV contribution and predictable
within perturbative QCD, while the $\LQ^3\,r^2$
potential is IR dominated and unpredictable in perturbative QCD.
It can be clarified by the combined analyses of IR renormalons,
OPE in Wilsonian EFT, and 
separation of IR contributions by
log resummations and  contour integrals.
\begin{itemize}
\item[(2)] 
Necessity of renormalization of Wilson coefficients at IR.
\end{itemize}
IR renormalization of
Wilson coefficients needs to be performed {\it by hand},
in a modern framework of
EFT which uses scale separation in
dimensional regularization.
This is because, identification of contributions from the
$\LQ$ scale is not automatic at each order of perturbative expansion.
One could inflate uncertainties of predictions
if this prescription is not carried out properly.

In the application of perturbative QCD to the heavy quarkonium
states, it is intriguing to see mutual connections between
the running of $\alpha_s(q)$, the linear rise $\sim \LQ^2\,r$ of
the interquark potential, and the quark self-energy
of order $\LQ(\LQ/m)^{1/3}$ which resembles the
constituent quark mass.
The relations are made quantitative in the perturbative regime
of QCD, $r\simlt \LQ^{-1}$.
Since these are basic concepts which also appear in other fields of 
perturbative QCD, the relations may be useful.

\section{Concluding Remarks}

The first part of this lecture was devoted to an
overview of perturbative QCD from a modern viewpoint.
We explained the
theoretical formulation, in which
the following subjects were covered:
(a) The relation between purely perturbative 
predictions and predictions based on OPE of EFTs.
(b) Computational methods for higher-order corrections,
which also give solid foundations to EFTs.
(c) Roles of singularities in Feynman amplitudes therein.
(See Sec.~3.6 for the summary of the first part.)
We tried to give a unified view in terms of singularities in
amplitudes.

Contrastingly, in the current status it is difficult 
to present an overview of diverse physics phenomena described
by perturbative QCD from a unified viewpoint.
By the same token, 
when applying the above theoretical formulation to 
each of these phenomena, 
one needs to adapt the formulation by
incorporating specific features of the system of interest,
even though the basic concept is general.
For example, the relevant dynamical degrees of freedom
depend on the system and it happens that almost
as many varieties of evolution equations appear
as the number of different subjects of perturbative QCD.
The differences consist in which
dynamical variables are relevant and which variables
have been integrated out;
(in an ultimate form of the theory) it is hoped that
each of them can be cast into an EFT in general
as described in this lecture.
Analyses based on singularities in amplitudes
are useful in developing an EFT, after the procedure of its
construction is 
clarified or in the process of clarifying it. 

In the latter part of this lecture we reviewed 
an application of perturbative QCD to heavy quarkonium systems,
in particular to bottomonium states and the static QCD potential.
It is interesting that different theoretical frameworks
and various concepts are mutually linked and converge towards
a consistent picture.
Namely, IR renormalons in the purely perturbative computation,
OPE in an EFT, and separation of IR contributions after
log resummations, all point to a consistent result, which
also agrees with lattice results.
At the same time we find an interrelation between
the concepts of the running coupling constant,
linear potential and quark self-energies (which resemble
constituent quark masses) from a
microscopic viewpoint, although
the validity range is  restricted to the short-distance
region $r\simlt \LQ^{-1}$.
(See Sec.~4.4 for lessons drawn from the latter part.)
In principle, we expect that similar consistency checks
can be carried out in applications of perturbative QCD
to other physical systems.

As stated, we focused on singularities in Feynman amplitudes
to present a unified view.
As one carries out higher-loop computations explicitly, one realizes that
the computations are composed by series of processes of resolution of
singularities.
Using known algorithms\footnote{
For instance, the author uses combinations of the Laporta algorithm,
method of differential equation,
method of Mellin-Barnes integral representation,
reduction of multiple sums and integrals, and shuffle relations, etc.,
in analytic evaluations of higher-loop computations.
}
one resolves entangled singularities in loop integrals step
by step.
As a result, for instance, the structure of the singularities is encoded in
the characters ($\lambda_i$s) of generalized
MZVs in the final results of radiative corrections.
It is a goal of current researches to reveal the encoding mechanism
and as a consequence find an efficient 
way of computation.
In dimensional regularization separation of multiple scales
in a physical process is also determined by the structure of
singularities (analyticity) of each amplitude.
We can perform scale separation 
not only by using poles of propagators but also by 
using the singularity of order $\LQ$ generated by log
resummations.
We pointed out the usefulness of this method in
renormalizing Wilson coefficients
at IR.

Although barely covered in this lecture, presently 
rapid progress is made in the field related to LHC physics,
such as automatization of computations of radiative corrections
and developments of computational tools by
combining them with MC simulation programs.
It is expected that 
multitudes of computations
for high energy processes including radiative
corrections
will be achieved in near future.
We look forward to see how these developments
will lead to new physics insights.

A partial 
list of other future developments expected in perturbative QCD
is as follows.
\begin{itemize}
\item[(a)]
Predictions in collaboration with lattice computations.
They would be indispensable for perturbative QCD
to become a high precision science.
\item[(b)]
Understanding the essence of radiative corrections, such as
finding solutions to the questions raised in Sec.~3.5.
\item[(c)]
Systematic construction of EFTs (without recourse to
diagrammatic methods).
Since the present methods for constructing EFTs require
considerable expertise, for general education it is desirable to
develop a transparent method
in a field theoretical approach.
\item[(d)]
To establish a general prescription for resummation of logarithms and
separation of IR contributions for an arbitrary OPE.
\end{itemize}

Although perturbative QCD has made significant progress
over the years,
there exist certain observables whose perturbative
expansions do not exhibit
expected convergence behaviors,
for reasons not well understood.
It is necessary to investigate them in detail, where brute-force
computations of higher-order corrections would not suffice.
For example, it may be similar to the case of the 
static potential before canceling the leading renormalon 
in the $r$-independent constant. (See Fig.~\ref{PotUpto3Loops}.)
There are many studies to be done, and we would like to
see more flows towards unification of perturbative QCD.

\section*{Acknowledgements}

The author would like to dedicate this lecture  to the
late Prof.~Jiro Kodaira, who was leading the community of
perturbative QCD in Japan.
The author is grateful to the organizers of ``QCD Club,''
in particular to K.~Fukushima, for giving the opportunity of
this lecture.
This work was supported in part by Grant-in-Aid for
scientific research No.\ 23540281 from
MEXT, Japan.

\vspace*{15mm}

\appendix
\clfn
\section*{Appendix: Proof of Eq.~(\ref{ImportantRel})}
\label{appA}

We give a proof of Eq.~(\ref{ImportantRel}):
\bea
V_{\rm UV}(r;\mu_f) - \Bigl[ V_C(r) + \sigma r \Bigr] = 
{\rm const.}+{\cal O}(\mu_f^3 r^2) .
\label{ImportantRel2}
\eea

\begin{figure}[t]\centering
\includegraphics[width=16cm]{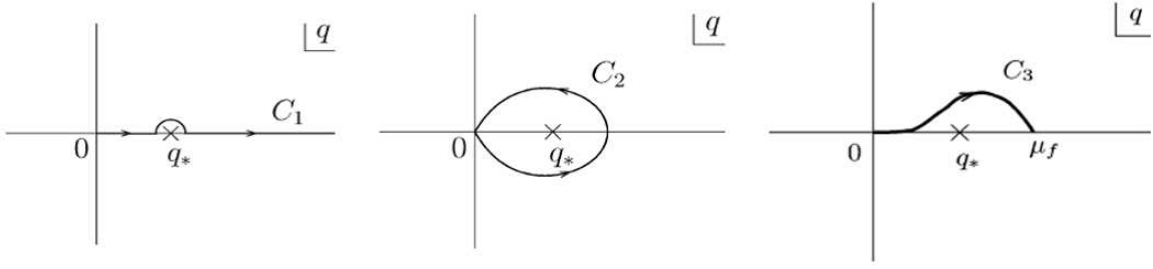}
\caption{\small
Contours $C_1$, $C_2$ and $C_3$ of integrals in
Eq.~(\ref{proofdiff1}).
\label{AllContours}
}
\end{figure}

$V_{\rm UV}(r)$ is defined in Eq.~(\ref{VUV}).
After integrating over the angular variables, it can be expressed
as a one-parameter integral in the range $\mu_f<q<\infty$:
\bea
V_{\rm UV}(r;\mu_f) = 
- \frac{2C_F}{\pi} 
\int_{\mu_f}^\infty\! dq \, \frac{\sin{qr}}{qr} \, \alpha_V(q) 
=- \frac{2C_F}{\pi} \, {\rm Im}
\int_{\mu_f}^\infty\! dq \, \frac{e^{iqr}}{qr} \, \alpha_V(q) .
\eea
The formulas for ${V_C(r)}$ and ${\sigma\,r}$ are given as contour
integrals in the complex $q$-plane by
Eqs.~(\ref{formula-Vc}) and (\ref{formula-sigma}).
Hence, the difference can be written as
\bea
V_{\rm UV}(r;\mu_f) - \Bigl[ V_C(r) + \sigma r \Bigr] = 
\frac{C_F}{\pi i}
\int_{C_2}\! dq \, \frac{\alpha_V(q) }{qr}
+ \frac{2C_F}{\pi} \, {\rm Im}
\int_{C_3}\! dq \, \frac{e^{iqr}}{qr} \, \alpha_V(q) -\sigma r .
\label{proofdiff1}
\eea
See Fig.~\ref{AllContours} for the contours $C_1$, $C_2$, $C_3$.
Since $\mu_f r\ll 1$, along the contour $C_3$ we can expand
the Fourier factor as
$
e^{iqr}=1+iqr+\frac{1}{2}(iqr)^2+\cdots
$.
Then we can rewrite the first and third terms as
\bea
&&
\frac{2C_F}{\pi } \, {\rm Im}
\int_{C_3}\! dq \, \frac{\alpha_V(q) }{qr}
=
-\frac{C_F}{\pi i}
\int_{C_2}\! dq \, \frac{\alpha_V(q) }{qr} ,
\label{vanishmuf1}
\\&&
\frac{2C_F}{\pi } \, {\rm Im}
\int_{C_3}\! dq \, \bigl(-\frac{1}{2}qr\bigr)\alpha_V(q) 
=
-\frac{C_F}{\pi i}
\int_{C_2}\! dq \, \Bigl(-\frac{1}{2}qr\Bigr)\alpha_V(q) .
\label{vanishmuf2}
\eea
These terms are canceled and the rest has a form\,
${\rm const.}+{\cal O}(\mu_f^3 r^2)$, which completes the proof.
It is in the above equations, when the contour $C_3$ is
changed to $C_2$, 
that $\mu_f$ dependences of the ``Coulomb''+linear terms are shown
to vanish.
The same change of contour does not apply to the constant and
$r^2$ terms (due to existence of $i$), hence the $\mu_f$ dependence
remain in these terms.

In essence, the above proof can be interpreted as follows.
IR contributions in
$V_{\rm QCD}(r)$ can be subtracted
as contour integrals along $C_3$ near the IR
singularity $q=q_*$ of $\alpha_V(q)$.
It is possible to decompose $V_{\rm UV}(r;\mu_f)$ 
into the form close to 
Eq.~(\ref{naive-expansion}).
The parts which are sensitive to the IR cut-off $\mu_f$
are the parts which can be absorbed into the pole
mass or the non-perturbative
matrix elements.
${V_C(r)}$ and ${\sigma\,r}$
have forms genuinely UV; it
is consistent since there are no IR operators which
can absorb the IR part of these $r$ dependences.


\begin{thebibliography}{99}

\bibitem{KodairaNote}
J.~Kodaira, 
``QCD: Past, Present and Future,'' based on the talk
in Symposium {\it ``30 years of QCD''}
at Japan Physics Society Meeting,
at Tokyo University of Science, Noda, Mar.\ 2005,
\texttt{http://www.jahep.org/hepnews/2005/
Vol24No2-2005.7.8.9kodaira.pdf}
(in Japanese).

\bibitem{Mueller:1989hs}
  ``Perturbative Quantum Chromodynamics,'' ed.\ A.~H.~Mueller,
  (World Scientific, Singapore, 1989).

\bibitem{Ellis:1991qj}
  R.~K.~Ellis, W.~J.~Stirling and B.~R.~Webber,
  ``QCD and collider physics,''
  Camb.\ Monogr.\ Part.\ Phys.\ Nucl.\ Phys.\ Cosmol.\  {\bf 8} (1996) 1.



\bibitem{Collins:2011zzd}
  J.~Collins,
  ``Foundations of perturbative QCD,''
  (Cambridge monographs on particle physics, nuclear physics and cosmology. 32).

\bibitem{Manohar:2000dt}
  A.~V.~Manohar and M.~B.~Wise,
  ``Heavy quark physics,''
  Camb.\ Monogr.\ Part.\ Phys.\ Nucl.\ Phys.\ Cosmol.\  {\bf 10} (2000) 1.

\bibitem{Brambilla:2004jw}
  N.~Brambilla, A.~Pineda, J.~Soto and A.~Vairo,
  ``Effective field theories for heavy quarkonium,''
  Rev.\ Mod.\ Phys.\  {\bf 77} (2005) 1423
  [hep-ph/0410047].

\bibitem{Smirnov:2004ym}
  V.~A.~Smirnov,
  ``Evaluating Feynman integrals,''
  Springer Tracts Mod.\ Phys.\  {\bf 211} (2004) 1.



\bibitem{Collins:1989gx}
  J.~C.~Collins, D.~E.~Soper and G.~F.~Sterman,
  Adv.\ Ser.\ Direct.\ High Energy Phys.\  {\bf 5} (1988) 1
  [hep-ph/0409313].
  (The first chapter of \cite{Mueller:1989hs}.)

\bibitem{Brambilla:2004wf}
  N.~Brambilla {\it et al.}  [Quarkonium Working Group Collaboration],
  hep-ph/0412158.

\bibitem{Brambilla:2010cs}
  N.~Brambilla, S.~Eidelman, B.~K.~Heltsley, R.~Vogt, G.~T.~Bodwin, E.~Eichten, A.~D.~Frawley and A.~B.~Meyer {\it et al.},
  Eur.\ Phys.\ J.\ C {\bf 71} (2011) 1534
  [arXiv:1010.5827 [hep-ph]].

\bibitem{Bogner:2007cr}
  C.~Bogner and S.~Weinzierl,
  Comput.\ Phys.\ Commun.\  {\bf 178} (2008) 596
  [arXiv:0709.4092 [hep-ph]].

\bibitem{Binoth:2000ps}
  T.~Binoth and G.~Heinrich,
  Nucl.\ Phys.\ B {\bf 585} (2000) 741
  [hep-ph/0004013].
  
\bibitem{Binoth:2003ak}
  T.~Binoth and G.~Heinrich,
  Nucl.\ Phys.\ B {\bf 680} (2004) 375
  [hep-ph/0305234].

\bibitem{Smirnov:2002pj}
  V.~A.~Smirnov,
  ``Applied asymptotic expansions in momenta and masses,''
  Springer Tracts Mod.\ Phys.\  {\bf 177} (2002) 1.

\bibitem{Beneke:1998ui}
  M.~Beneke,
  Phys.\ Rept.\  {\bf 317} (1999) 1
  [hep-ph/9807443].

\bibitem{Beneke:1994qe}
  M.~Beneke and V.~M.~Braun,
  Phys.\ Lett.\ B {\bf 348} (1995) 513
  [hep-ph/9411229].

\bibitem{Chetyrkin:1981qh}
  K.~G.~Chetyrkin and F.~V.~Tkachov,
  Nucl.\ Phys.\  B {\bf 192}, 159 (1981).

\bibitem{Jantzen:2011nz}
  B.~Jantzen,
  JHEP {\bf 1112} (2011) 076
  [arXiv:1111.2589 [hep-ph]].

\bibitem{Laporta:1996mq}
  S.~Laporta and E.~Remiddi,
  Phys.\ Lett.\ B {\bf 379} (1996) 283
  [hep-ph/9602417].

\bibitem{Aoyama:2012wj}
  T.~Aoyama, M.~Hayakawa, T.~Kinoshita and M.~Nio,
  Phys.\ Rev.\ Lett.\  {\bf 109} (2012) 111807
  [arXiv:1205.5368 [hep-ph]].

\bibitem{Thm1}
A.~B.~Goncharov, {\it Multiple polylogarithms and mixed
Tate motives}, arxiv:math.AG/0103059.

\bibitem{Thm2}
T.~Terasoma, {\it Mixed Tate motives and multiple zeta values},
Invent.\ Math.\ {\bf 149} (2002) 339.

\bibitem{Thm3}
P.~Deligne and A.~Goncharov, 
{\it Groupes fondamentaux motiviques de Tate mixte},
Ann.\ Sci.\ Ecole Norm.\ Sup.\ (4) {\bf 38} (2005) 1.

\bibitem{ShuffleRel}
K.~Ihara, M.~Kaneko, and D.~Zagier, 
{\it Derivation and double shuffle relations for multiple
zeta values}, Compos.\ Math., {\bf 142} (2006) 307.

\bibitem{Anzai:2012xw}
  C.~Anzai and Y.~Sumino,
  J.\ Math.\ Phys.\ {\bf 54} (2013) 033514
  [arXiv:1211.5204 [hep-th]].

\bibitem{Broadhurst:1996az}
  D.~J.~Broadhurst,
  hep-th/9604128.

\bibitem{Broadhurst:1998rz}
  D.~J.~Broadhurst,
  Eur.\ Phys.\ J.\ C {\bf 8} (1999) 311
  [hep-th/9803091].
\bibitem{Brown:2010bw}
  F.~Brown and O.~Schnetz,
  arXiv:1006.4064 [math.AG].

\bibitem{Remiddi:1997ny}
  E.~Remiddi,
  Nuovo Cim.\ A {\bf 110} (1997) 1435
  [hep-th/9711188].

\bibitem{ATLASdata}
ATLAS Group,
\texttt{https://twiki.cern.ch/twiki/bin/view/AtlasPublic/
\\
StandardModelPublicCollisionPlots.
}

\bibitem{Beringer:1900zz}
  J.~Beringer {\it et al.}  [Particle Data Group Collaboration],
  Phys.\ Rev.\ D {\bf 86} (2012) 010001.

\bibitem{ATLAS:2014wva}
  [ATLAS and CDF and CMS and D0 Collaborations],
  arXiv:1403.4427 [hep-ex].

\bibitem{Anzai:2009tm} 
  C.~Anzai, Y.~Kiyo and Y.~Sumino,
Phys.\ Rev.\ Lett.\  {\bf 104}, 112003 (2010).  

\bibitem{Smirnov:2009fh} 
  A.~V.~Smirnov, V.~A.~Smirnov and M.~Steinhauser,
Phys.\ Rev.\ Lett.\  {\bf 104}, 112002 (2010).  


\bibitem{Necco:2001xg}
  S.~Necco and R.~Sommer,
  Nucl.\ Phys.\ B {\bf 622}, 328 (2002).

\bibitem{Takahashi:2002bw}
  T.~T.~Takahashi, H.~Suganuma, Y.~Nemoto and H.~Matsufuru,
  Phys.\ Rev.\ D {\bf 65}, 114509 (2002).

\bibitem{Aoki:2002uc}
  S.~Aoki {\it et al.}  [JLQCD Collaboration],
  Phys.\ Rev.\ D {\bf 68}, 054502 (2003).

\bibitem{Aglietti:1995tg}
U.~Aglietti and Z.~Ligeti,
Phys.\ Lett.\ B {\bf 364}, 75 (1995).

\bibitem{Pineda:id}
A.~Pineda, Ph.D. Thesis,
\\
\texttt{http://www.slac.stanford.edu/spires/find/hep/www?irn=5399084}.

\bibitem{Hoang:1998nz}
A.~H.~Hoang, M.~C.~Smith, T.~Stelzer and S.~Willenbrock,
Phys.\ Rev.\ D {\bf 59}, 114014 (1999).

\bibitem{Beneke:1998rk}
M.~Beneke,
Phys.\ Lett.\ B {\bf 434}, 115 (1998).

\bibitem{Chetyrkin:1997sg}
  K.~G.~Chetyrkin, B.~A.~Kniehl and M.~Steinhauser,
  Phys.\ Rev.\ Lett.\  {\bf 79} (1997) 2184
  [hep-ph/9706430].

\bibitem{Sumino:2001eh}
Y.~Sumino,
Phys.\ Rev.\ D {\bf 65}, 054003 (2002).

\bibitem{Brambilla:1999qa}
N.~Brambilla, A.~Pineda, J.~Soto and A.~Vairo,
Phys.\ Rev.\ D {\bf 60}, 091502 (1999).

\bibitem{Sumino:2004ht}
  Y.~Sumino,
  Phys.\ Lett.\ B {\bf 595} (2004) 387
  [hep-ph/0403242].

\bibitem{Sumino:2003yp}
  Y.~Sumino,
  Phys.\ Lett.\ B {\bf 571} (2003) 173
  [hep-ph/0303120].


\bibitem{Sumino:2005cq}
  Y.~Sumino,
  Phys.\ Rev.\  D {\bf 76}, 114009 (2007).

\bibitem{Pineda:2002se}
A.~Pineda,
J.\ Phys.\ G {\bf 29}, 371 (2003);

\bibitem{Kiyo:2013aea}
  Y.~Kiyo and Y.~Sumino,
  Phys.\ Lett.\ B {\bf 730} (2014) 76
  [arXiv:1309.6571 [hep-ph]].

\bibitem{Brambilla:2001fw}
  N.~Brambilla, Y.~Sumino and A.~Vairo,
  Phys.\ Lett.\ B {\bf 513} (2001) 381.



\end{thebibliography}
\end{document}